\documentclass[10pt,twocolumn,english,prd,superscriptaddress,nofootinbib,preprintnumbers,showpacs,floatfix]{revtex4-2}
\usepackage[T1]{fontenc}
\usepackage[utf8]{inputenc}
\usepackage{lmodern}
\usepackage{amsmath,amssymb,amsthm,physics,mathrsfs}
\usepackage{graphicx,epstopdf}
\usepackage{booktabs,adjustbox}
\usepackage[dvipsnames]{xcolor}
\usepackage{textgreek}
\usepackage{orcidlink}
\usepackage{hyperref}
\hypersetup{ colorlinks=true, linkcolor=blue, citecolor=cyan, urlcolor=blue}

\allowdisplaybreaks

\newcommand{\ABTU}{Department of Physics, Al-Hussein Bin Talal University, 71111, Ma'an, Jordan}

\begin{document}

\title{Quasi-periodic oscillations in Letelier  black holes within Kalb-Ramond gravity embedded in a uniform external magnetic field}

\author{Ahmad Al-Badawi\orcidlink{0000-0002-3127-3453}}
\email[Email: ]{ahmadbadawi@ahu.edu.jo }
\affiliation{\ABTU}

\author{Shokhzod Jumaniyozov \orcidlink{0009-0009-6254-5608}}
\email[Email: ]{sh.jumaniyozov@newuu.uz}
\affiliation{Kimyo International University in Tashkent, Shota Rustaveli Street 156, Tashkent 100121, Uzbekistan}
\affiliation{Tashkent State Technical University, Tashkent 100095, Uzbekistan}

\date{\today}

\begin{abstract}
In the present work, we study the motion of charged test particles around the Letelier black hole (a black hole surrounded by a cloud of strings, CoS) in Kalb--Ramond (KR) gravity, which is immersed in an external asymptotically uniform magnetic field directed along the symmetry axis. The metric contains the CoS parameter $\alpha$ and the Lorentz-violating KR parameter $\ell$ only in the combination $A=(1-\alpha)/(1-\ell)$. Thus the two effects can cancel each other, and for $\alpha=\ell$ the Schwarzschild metric is recovered. The spacetime is not Ricci flat, and for this reason the Wald potential $A_\phi\propto r^2\sin^2\theta$ does not satisfy the Maxwell equations. We obtain the regular source-free magnetic test field in closed form in terms of the Gauss hypergeometric function, $A_\phi=\tfrac{B}{2}\Psi(r)\sin^2\theta$ with $\Psi\propto{}_2F_1(s_+-1,-s_+;1;1-Ar/2M)$ and $s_+=\tfrac12\big(1+\sqrt{1+8/A}\big)$; for $A=1$ it reduces to $\Psi=r^2$. With this field we derive the equations of motion and the effective potential, and analyze circular orbits, the innermost stable circular orbit (ISCO) and some typical escaping, captured and curled trajectories. We also obtain closed-form expressions for the orbital, radial and vertical epicyclic frequencies and check them by direct numerical integration of the full equations of motion. Then, in the framework of the relativistic precession model, we carry out a Markov chain Monte Carlo (MCMC) analysis of the twin-peak high-frequency quasi-periodic oscillations (QPOs) observed in GRO J1655$-$40, XTE J1550$-$564 and GRS 1915$+$105 for four nested models. Nested sampling is used in addition to calculate the Bayesian evidences and the weights of the two (Larmor and anti-Larmor) magnetic modes of the posterior. We find that the Schwarzschild model is excluded for GRO J1655$-$40 and XTE J1550$-$564 ($\Delta{\rm AIC}\simeq110$). The purely geometric deformation can fit these two sources only for $A\simeq1.18$--$1.21$, and then $\ell$ goes to the upper bound allowed by current KR constraints. On the other hand, a weak Larmor-type magnetic coupling $\beta=qBM/(2m)\simeq-0.02$ fits both sources at their dynamical masses and is preferred by the evidence ($\ln B=3.2$ and $6.1$). For GRS 1915$+$105 all models, the Schwarzschild one included, are statistically equivalent. Since the QPO data constrain only $A$, the CoS and the KR field stay degenerate along $\alpha\simeq\ell$, and other observables are necessary to break this degeneracy.
\end{abstract}

\maketitle

\section{Introduction}\label{sec1}

Lorentz invariance is one of the basic principles of general relativity (GR) and of quantum field theory. However, several approaches to quantum gravity suggest that it may be only an approximate symmetry which emerges at low energies. Spontaneous breaking of Lorentz symmetry was first noticed in string field theory, where tensor fields can have non-vanishing vacuum expectation values~\cite{Kostelecky:1988zi}. The Standard-Model Extension gives an effective-field-theory framework for describing such effects both in the matter and in the gravitational sectors~\cite{Colladay:1998fq,Kostelecky:2003fs,Bluhm:2004ep}, and many experiments have been carried out to search for possible Lorentz-violating signals~\cite{Mattingly:2005re}. In gravity, the simplest examples are the bumblebee models, in which a vector field gets a vacuum expectation value; these models have exact Schwarzschild-like black-hole solutions~\cite{Casana:2017jkc}.

Another candidate, which is in some respects more natural, is the Kalb--Ramond (KR) field. It is a rank-two antisymmetric tensor $B_{\mu\nu}$ appearing in the bosonic sector of closed-string theory~\cite{Kalb:1974yc}. If the KR field is non-minimally coupled to gravity and has a non-zero vacuum expectation value, Lorentz symmetry is spontaneously broken. In this case static and spherically symmetric black-hole solutions differ from the Schwarzschild one by a dimensionless Lorentz-violating parameter $\ell$~\cite{Lessa:2019bgi,Yang:2023wtu,Duan:2023gng}. Such solutions have been tested using Solar-System data, black-hole shadows and lensing~\cite{Junior:2024ety,Junior:2024vdk}, periodic orbits and gravitational waveforms~\cite{Junior:2024tmi}, tidal and Doppler effects~\cite{Cordeiro:2025cfo}, as well as the dynamics of neutral, charged, spinning and magnetized particles and their quasi-periodic oscillations~\cite{Jumaniyozov:2024eah,Jumaniyozov:2025wcs,Jumaniyozov:2025dyy,Jumaniyozov:2025lox,Jumaniyozov:2025nnj}. In Ref.~\cite{Junior:2024ety} the range $-0.70\lesssim\ell\lesssim0.19$ was obtained from observations.

In reality, black holes are not isolated objects. A simple model of the matter around a black hole is the cloud of strings (CoS) proposed by Letelier~\cite{Letelier:1979ej}. In this model a collection of one-dimensional string-like objects produces an anisotropic energy--momentum tensor and a solid-angle deficit, similar to the case of the global monopole~\cite{Barriola:1989hx}. Black holes with a CoS have been studied in many modified theories~\cite{Jumaniyozov:2026gxv,Belchior:2025xam}. Recently, the KR black hole surrounded by a CoS (below we call it the Letelier--KR black hole) was obtained and tested with EHT shadow and X-ray QPO data~\cite{AHMED2026102315,Ahmed:2025oxp}. The metric function of this solution contains the CoS parameter $\alpha$ and the KR parameter $\ell$ only in the combination $(1-\alpha)/(1-\ell)$. For this reason it is a good example to see which effects can be separated by observations and which cannot.

Magnetic fields are also an important part of the black hole environment. After the well-known work of Wald~\cite{Wald:1974np}, it has been understood that black holes in magnetic fields play a key role in jet formation and energy extraction~\cite{Blandford:1977ds}. Polarimetric images of M87* show that ordered magnetic fields exist near the horizon~\cite{EventHorizonTelescope:2021srq}. Even a weak magnetic field has a strong influence on the motion of charged particles: it shifts the ISCO and the epicyclic frequencies and leads to chaotic and curled trajectories~\cite{Aliev:2002nw,Frolov:2010mi,Kolos:2015iva,Stuchlik:2015nlt,Tursunov:2016dss,Jumaniyozov:2026lbf}. Here one technical point should be noted. The Wald method is based on the fact that Killing vectors solve the source-free Maxwell equations, but this is true only in Ricci-flat spacetimes. In the KR and Letelier--KR geometries the Wald potential $A_\phi=\frac{B}{2}r^2\sin^2\theta$ is \emph{not} a solution of the Maxwell equations, and the magnetic test field has to be found directly from $\nabla_\mu F^{\mu\nu}=0$, as was pointed out recently in Ref.~\cite{Ahmed:2026hmy}.

Quasi-periodic oscillations (QPOs) in the X-ray flux of accreting black holes are a useful tool for testing strong gravity~\cite{Psaltis:2008bb,Ingram:2019mna}. Twin-peak high-frequency QPOs have been observed in the microquasars GRO J1655$-$40, XTE J1550$-$564 and GRS 1915$+$105~\cite{Strohmayer:2001yn,Remillard:2002cy,Remillard:2006fc,Belloni:2012sv}. Their frequencies are close to the orbital and epicyclic frequencies of particles near the ISCO, and phenomenological models, for example the relativistic precession model (RPM)~\cite{Stella:1997tc,Stella:1999sj,Motta:2013wga} and the resonance models~\cite{Abramowicz:2001bi}, connect them with test-particle motion. Using Markov chain Monte Carlo (MCMC) methods, QPO data have been applied to constrain many modified black-hole spacetimes, among them the KR black hole in an external magnetic field~\cite{Rodrigues:2026ofx,Ahmed:2026hmy}.

In this paper, we consider the Letelier--KR black hole in an external asymptotically uniform magnetic field directed along the $z$-axis. First, we show that the regular magnetic test field in this spacetime can be written in closed form through a Gauss hypergeometric function, and that it coincides with the Wald profile when $\alpha=\ell$. Then we obtain closed-form expressions for the circular orbits and for the three fundamental frequencies of charged particles, and check them by direct numerical integration of the equations of motion. We also analyze the effective potential, the ISCO and several typical escaping, captured and curled trajectories. Finally, we compare the model with the QPO data of three microquasars using an MCMC analysis of four nested models, and we use information criteria to see which ingredients are really needed to describe the data.

The paper is organized as follows. In Sec.~\ref{sec2} we present the metric and find the magnetic test field. The equations of motion, the effective potential and the circular orbits are derived in Sec.~\ref{sec3}, and the trajectories of charged particles are discussed in Sec.~\ref{sec4}. In Sec.~\ref{sec5} we study the fundamental frequencies and perform the MCMC analysis of the QPO data. Our conclusions are given in Sec.~\ref{sec6}, and some technical details are presented in the Appendix. Throughout the paper we use geometrized units $G=c=1$ and the signature $(-,+,+,+)$; for the QPO frequencies we return to physical units.

\section{Letelier Kalb-Ramond black hole and the magnetic test field}\label{sec2}

\subsection{Spacetime geometry}

The line element of the Letelier black hole in KR gravity~\cite{AHMED2026102315} has the form
\begin{equation}
	ds^{2}=-f(r)dt^{2}+\frac{dr^{2}}{f(r)}+r^{2}\left(d{\theta}^{2}+\sin^{2}{\theta}\,d{\phi}^{2}\right),
	\label{ds2}
\end{equation}
with
\begin{equation}
	f(r)=\frac{1-\alpha}{1-\ell}-\frac{2M}{r}\equiv A-\frac{2M}{r},\qquad A\equiv\frac{1-\alpha}{1-\ell}.
	\label{eq:metricFunction}
\end{equation}
where $M$ is the mass parameter, $\ell$ is the Lorentz-violating parameter which comes from the vacuum expectation value of the KR field, and $\alpha$ is an integration constant connected with the energy density of the CoS ($0\le\alpha<1$). In the limit $\alpha\to0$ we get the Schwarzschild-like KR solution $f=1/(1-\ell)-2M/r$~\cite{Yang:2023wtu,Junior:2024ety}, and for $\ell\to0$ we get the Letelier black hole of GR with $f=1-\alpha-2M/r$~\cite{Letelier:1979ej}. The event horizon radius is
\begin{equation}
	r_{\rm h}=\frac{2M}{A}=\frac{2M(1-\ell)}{1-\alpha}.
	\label{eq:horizon}
\end{equation}

There are two properties of Eq.~\eqref{eq:metricFunction} which are important for the rest of the paper. The first one is that \emph{$\alpha$ and $\ell$ enter the metric only through $A$}. The CoS ($\alpha>0$) makes $A$ smaller, while a positive KR parameter makes it larger, and when $\alpha=\ell$ these two effects cancel and we come back to the Schwarzschild line element. Therefore, every observable related to test-particle motion in this spacetime depends only on $(M,A)$, and observations can constrain at most the combination $A$. The second property is that for $A\neq1$ the spacetime is not asymptotically flat. After the change $\tilde t=\sqrt{A}\,t$ and $\tilde r=r/\sqrt{A}$, the line element at large distances takes the form $ds^2\to-d\tilde t^2+d\tilde r^2+A\,\tilde r^2d\Omega^2$. This is flat space with a solid-angle excess ($A>1$) or deficit ($A<1$), like in the global monopole case~\cite{Barriola:1989hx}. As a result, a static observer far from the black hole measures the time
\begin{equation}
	t_{\rm phys}=\sqrt{A}\,t ,
	\label{eq:tphys}
\end{equation}
and the frequencies calculated with respect to the coordinate time $t$ should be divided by $\sqrt{A}$ before comparing them with observational data. For $\alpha=0$ this normalization is the same as the one used for the KR black hole in Ref.~\cite{Ahmed:2026hmy}.

\begin{figure*}[t]
	\centering
	\includegraphics[width=0.9\textwidth]{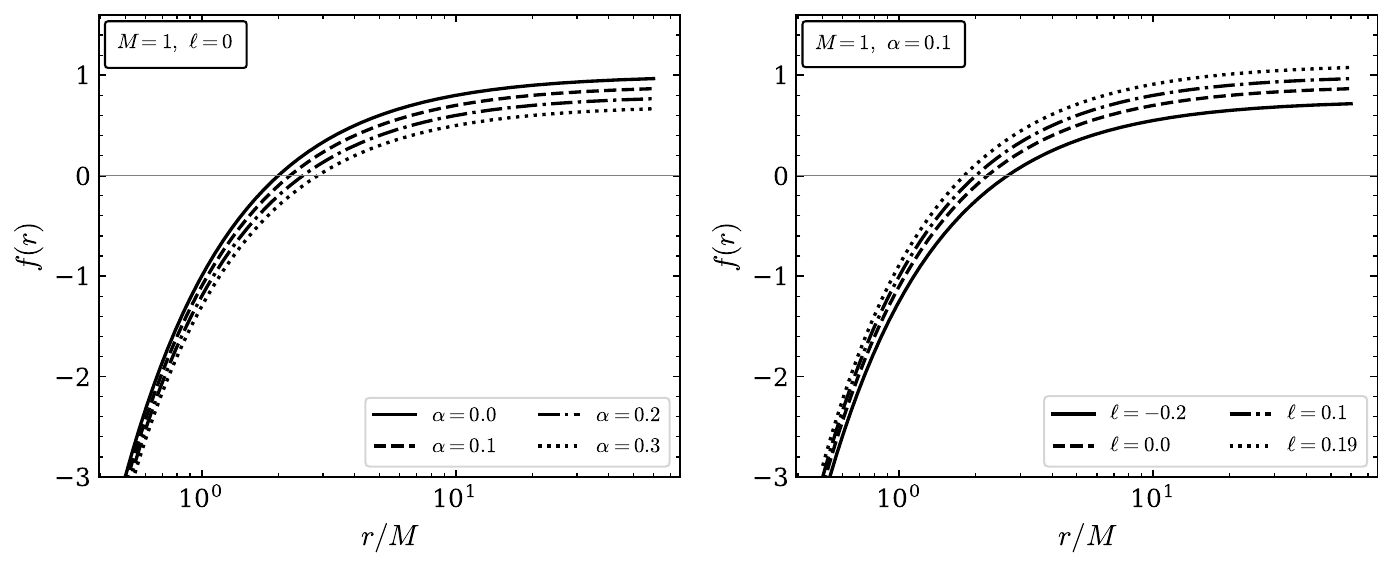}
	\caption{Metric function $f(r)$ of the Letelier--KR black hole for $M=1$. Left: different values of the CoS parameter $\alpha$ at $\ell=0$. Right: different values of the KR parameter $\ell$ at $\alpha=0.1$. The zero crossings give the event horizon $r_{\rm h}=2M/A$, while the large-$r$ value $f\to A$ fixes the asymptotic normalization of time.}
	\label{fig:metric}
\end{figure*}

\begin{figure*}[t]
	\centering
	\includegraphics[width=0.9\textwidth]{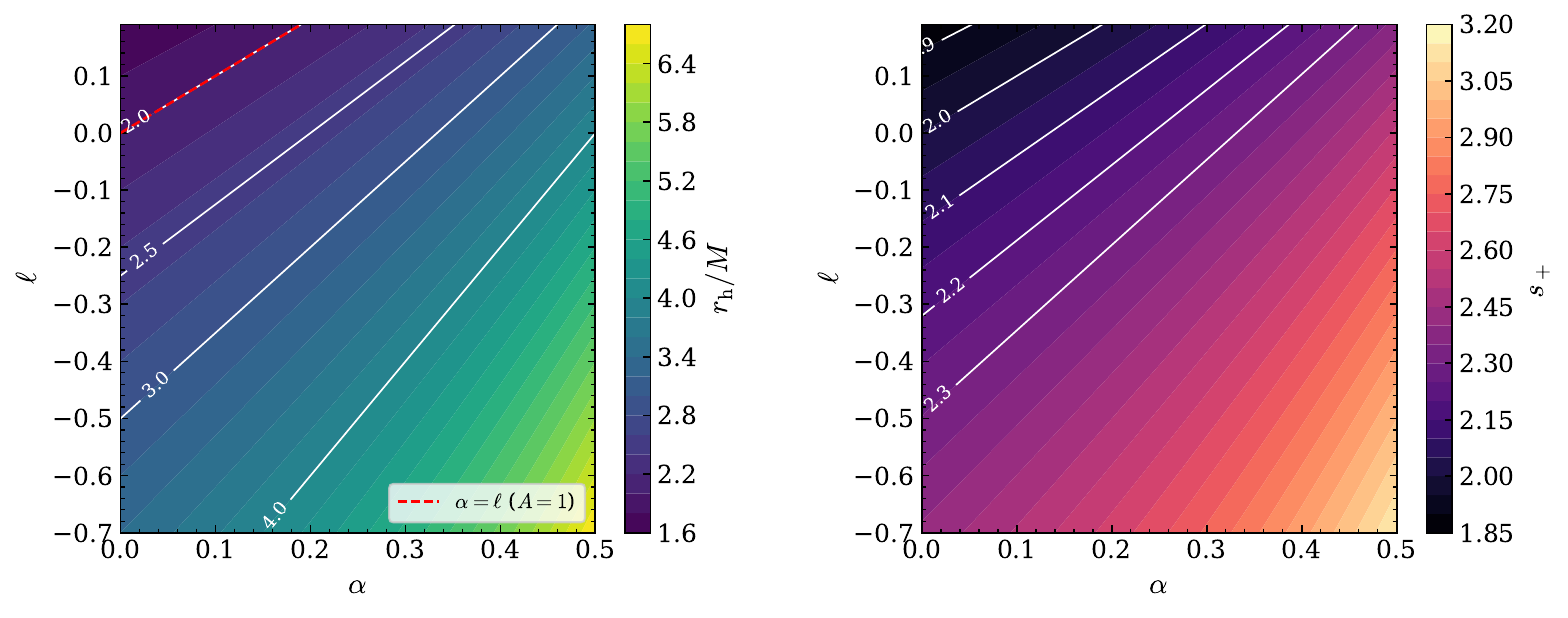}
	\caption{Left: event-horizon radius $r_{\rm h}/M$ in the $(\alpha,\ell)$ plane; the red dashed line $\alpha=\ell$ ($A=1$) corresponds to the Schwarzschild geometry. Right: exponent $s_+$ of the growing branch of the magnetic profile, $\Psi\sim r^{s_+}$, Eq.~\eqref{eq:spm}; $s_+=2$ (uniform Wald field) only for $A=1$.}
	\label{fig:horizon}
\end{figure*}

In Fig.~\ref{fig:metric} we plot the metric function for several values of $\alpha$ and $\ell$, and the left panel of Fig.~\ref{fig:horizon} shows the horizon radius in the $(\alpha,\ell)$ plane. One can see that the CoS moves the horizon outward and decreases the asymptotic value of $f$, and $\ell>0$ does the opposite. The dashed line $\alpha=\ell$ in Fig.~\ref{fig:horizon} corresponds to the configurations which are equivalent to Schwarzschild, with $r_{\rm h}=2M$.

\subsection{Source-free magnetic test field}

Now we consider an axisymmetric magnetic test field directed along the symmetry axis of the black hole. Its energy density is assumed to be small, so that it does not change the geometry. Wald showed that in Ricci-flat spacetimes the Killing vector $\xi_{(\phi)}=\partial_\phi$ is a solution of the source-free Maxwell equations, and for a field that is uniform at infinity this gives $A_\phi=\frac{B}{2}r^2\sin^2\theta$~\cite{Wald:1974np}. This result is based on the identity $\nabla_\nu\nabla^\nu\xi^\mu=-R^\mu{}_\nu\xi^\nu$, so it does not work when $R_{\mu\nu}\neq0$. This is exactly our case, because the CoS and the KR vacuum expectation value play the role of effective sources. For this reason we take
\begin{equation}
	A_\mu dx^\mu=\frac{B}{2}\Psi(r)\sin^2\theta\,d\phi ,
	\label{eq:Aphi}
\end{equation}
where the other components are zero, and find $\Psi(r)$ from the equation $\nabla_\mu F^{\mu\nu}=0$. The non-zero components of the field tensor are $F_{r\phi}=\frac{B}{2}\Psi'\sin^2\theta$ and $F_{\theta\phi}=B\Psi\sin\theta\cos\theta$. Taking into account $\sqrt{-g}=r^2\sin\theta$, from the $\nu=\phi$ component of the Maxwell equations we get
\begin{equation}
	\frac{d}{dr}\left[f(r)\frac{d\Psi}{dr}\right]-\frac{2\Psi}{r^2}=0,
	\label{eq:maxwell}
\end{equation}
and the other components are satisfied identically. At large $r$ we have $f\to A$, and Eq.~\eqref{eq:maxwell} has power-law solutions $\Psi\sim r^{s}$ with $As(s-1)=2$, which gives
\begin{equation}
	s_\pm=\frac12\left(1\pm\sqrt{1+\frac{8}{A}}\right).
	\label{eq:spm}
\end{equation}
In the case $A=1$ we have $s_+=2$, which is the uniform Wald field, and $s_-=-1$, which is a magnetic dipole.

It turns out that Eq.~\eqref{eq:maxwell} has an exact solution. In terms of the variable $z=r/r_{\rm h}=Ar/(2M)$ it takes the form $z(1-z)\Psi_{zz}-\Psi_z+(2/A)\Psi=0$. This is the hypergeometric equation with $a+b=-1$, $ab=-2/A$ and $c=-1$ (details are given in the Appendix). The solution regular at the event horizon reads
\begin{equation}
	\Psi(r)=\mathcal N\;{}_2F_1\!\left(s_+-1,\,-s_+;\,1;\,1-\frac{Ar}{2M}\right),
	\label{eq:Psi}
\end{equation}
The behaviour of this solution at large distances can be found from the standard connection formula for ${}_2F_1$ with large negative argument~\cite{Olver:2010dlmf}, ${}_2F_1\to C_+z^{s_+}+C_-z^{s_-}$, where $C_+=\Gamma(2s_+-1)/[\Gamma(s_+-1)\Gamma(s_++1)]$. The normalization constant is fixed by the condition
\begin{equation}
	\Psi(r)\simeq M^{2-s_+}r^{s_+},\qquad r\to\infty ,\qquad \mathcal N=\frac{M^{2-s_+}r_{\rm h}^{s_+}}{C_+},
	\label{eq:norm}
\end{equation}
so $B$ has the meaning of the amplitude of the field at large distances, in units suitable for this geometry. When $A=1$, we have $s_+=2$, $C_+=1$ and ${}_2F_1(1,-2;1;1-z)=z^2$, so Eq.~\eqref{eq:Psi} gives exactly the Wald profile $\Psi=r^2$. This is expected, since in this case the spacetime is Schwarzschild. The solution is regular at the horizon because the third parameter of the hypergeometric function in Eq.~\eqref{eq:Psi} is $c=1$, and one gets $\Psi'(r_{\rm h})=2\Psi(r_{\rm h})/[r_{\rm h}^2f'(r_{\rm h})]$. The derivatives can be calculated using $\frac{d}{dx}{}_2F_1(a,b;c;x)=\frac{ab}{c}{}_2F_1(a+1,b+1;c+1;x)$. To check Eq.~\eqref{eq:Psi}, we computed the residual of Eq.~\eqref{eq:maxwell} with 30-digit arithmetic (the relative residual is $\lesssim10^{-16}$ for $0.85\le A\le1.15$), and we also checked the asymptotic normalization numerically up to $r=10^5M$.

\begin{figure*}[t]
	\centering
	\includegraphics[width=\textwidth]{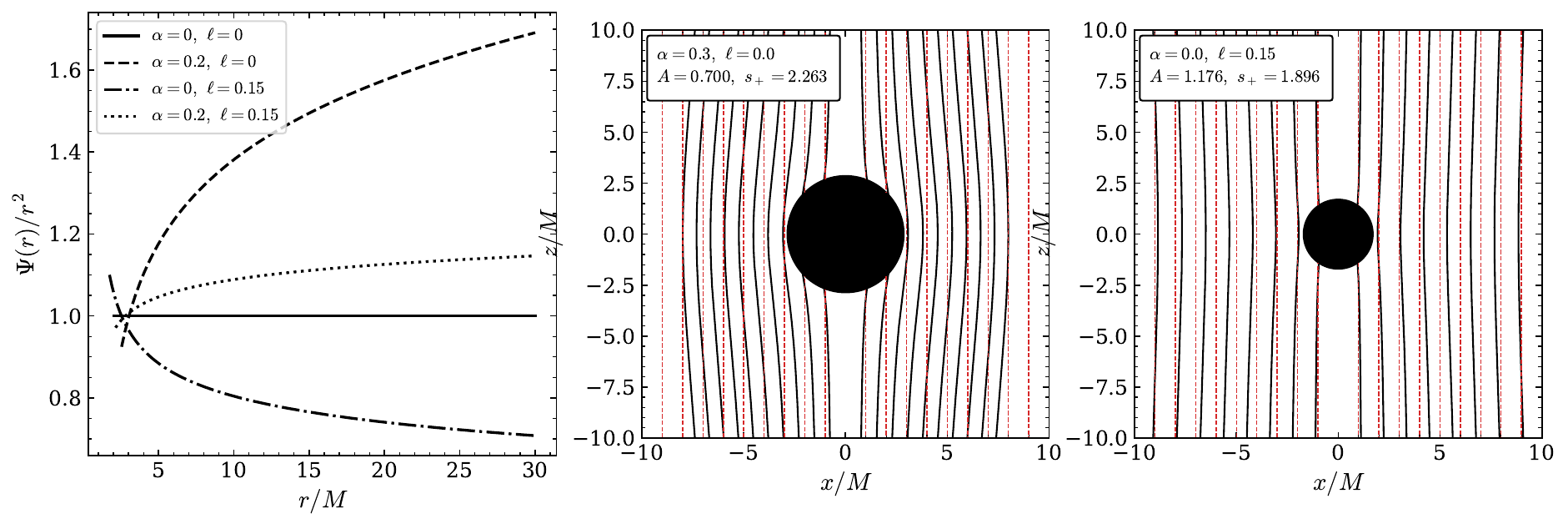}
	\caption{Left: ratio $\Psi(r)/r^2$ of the exact regular magnetic profile, Eq.~\eqref{eq:Psi}, to the Wald profile for several $(\alpha,\ell)$; the two coincide for $A=1$. Middle and right: magnetic field lines (level curves of $\Psi\sin^2\theta$) in the meridional plane for $(\alpha,\ell)=(0.3,0)$ and $(0,0.15)$ (solid black) compared with the Wald field lines $r^2\sin^2\theta={\rm const}$ (red dashed). The black disk is the event horizon.}
	\label{fig:magnetic}
\end{figure*}

For a static observer, the orthonormal components of the magnetic field are
\begin{equation}
	B^{\hat r}=\frac{B\,\Psi(r)}{r^2}\cos\theta ,\qquad B^{\hat\theta}=-\frac{B\sqrt{f(r)}}{2r}\,\Psi'(r)\sin\theta ,
	\label{eq:Bcomp}
\end{equation}
and for $A=1$ they become $B^{\hat r}=B\cos\theta$, $B^{\hat\theta}=-B\sqrt{f}\sin\theta$. At large distances the field slowly increases for $A<1$ and slowly decreases for $A>1$, as $|{\bf B}|\propto(r/M)^{s_+-2}$. In this sense we call the field \emph{asymptotically uniform-like}; it is the only regular source-free generalization of the Wald field to the Letelier--KR spacetime. Fig.~\ref{fig:magnetic} shows the ratio $\Psi/r^2$ and the magnetic field lines (level curves of $\Psi\sin^2\theta$) in comparison with the Wald case.

\section{Dynamics of charged particles}\label{sec3}

\subsection{Lagrangian, equations of motion and conserved quantities}

First, we consider the motion of a test particle with mass $m$ and charge $q$, which is described by the Lagrangian
\begin{align}
		\mathscr{L}&=\frac{m}{2}\,g_{\mu\nu}\dot x^\mu\dot x^\nu+qA_\mu\dot x^\mu\nonumber\\
		&=\frac{m}{2}\left[-f\dot{t}^{2}+\frac{\dot{r}^{2}}{f}+r^{2}\dot{\theta}^{2}+r^{2}\sin^{2}{\theta}\,\dot{\phi}^{2}\right]\nonumber\\
		&\quad+\frac{qB}{2}\Psi(r)\sin^{2}{\theta}\,\dot{\phi},
	\label{lagrangeana}
\end{align}
where the dot means derivative with respect to the proper time $\tau$. The Lagrangian does not depend on $t$ and $\phi$, and therefore there are two conserved quantities, the specific energy and the specific canonical angular momentum,
\begin{equation}
	\mathcal{E}=f(r)\,\dot t ,\qquad
	\mathcal{L}=r^2\sin^2\theta\,\dot\phi+\beta\,\Psi(r)\sin^2\theta ,
	\label{eq:EL}
\end{equation}
Here we introduced the magnetic coupling parameter
\begin{equation}
	\beta=\frac{qB}{2m},
	\label{eq:beta}
\end{equation}
which has the dimension of inverse length. In what follows we put $M=1$, and then $\beta$ means the dimensionless quantity $qBM/(2m)$ (in physical units $\beta=qBGM/(2mc^3)$). Depending on the sign of $\beta$ with respect to $\mathcal L$, there are two different physical situations~\cite{Kolos:2015iva,Tursunov:2016dss}. If $\mathcal L>0$ and $\beta>0$, the Lorentz force is repulsive (anti-Larmor motion), and if $\beta<0$, it is attractive (Larmor motion). Since the equations of motion do not change under $(\mathcal L,\beta)\to(-\mathcal L,-\beta)$, we can take $\mathcal L>0$ without loss of generality. Then Eq.~\eqref{eq:EL} gives
\begin{equation}
	\dot t=\frac{\mathcal E}{f(r)},\qquad
	\dot\phi=\frac{\mathcal L-\beta\Psi(r)\sin^2\theta}{r^2\sin^2\theta}.
	\label{eq:tphidot}
\end{equation}
When $A=1$, we have $\Psi=r^2$, and the second expression becomes $\dot\phi=\mathcal L/(r^2\sin^2\theta)-\beta$. This is the well-known result for the Schwarzschild black hole in a uniform magnetic field~\cite{Frolov:2010mi,Kolos:2015iva,Rodrigues:2026ofx}. The other Euler--Lagrange equations are
\begin{align}
	\ddot{\theta}&=-\frac{2}{r}\dot{r}\dot{\theta}+\frac{\cos\theta}{r^4\sin^3\theta}\,\mathcal X\left(\mathcal X+2\beta\Psi\sin^2\theta\right),
	\label{eq:thetaddot}\\
	\ddot{r}&=\frac{f'}{2f}\dot{r}^{2}-\frac{f'\mathcal{E}^{2}}{2f}+rf\dot{\theta}^{2}
	+\frac{f\,\mathcal X}{r^{2}\sin^{2}\theta}\left(\frac{\mathcal X}{r}+\beta\Psi'\sin^2\theta\right),
	\label{eq:rddot}
\end{align}
where $\mathcal X(r,\theta)\equiv\mathcal L-\beta\Psi(r)\sin^2\theta=r^2\sin^2\theta\,\dot\phi$ is the mechanical (kinetic) angular momentum. In numerical calculations it is easier to work with the equivalent Hamiltonian
\begin{equation}
	H=\frac12\left[f p_r^2+\frac{p_\theta^2}{r^2}+\frac{\mathcal X^2}{r^2\sin^2\theta}-\frac{\mathcal E^2}{f}+1\right]=0 ,
	\label{eq:Ham}
\end{equation}
The condition $H=0$ follows from the normalization $g_{\mu\nu}\dot x^\mu\dot x^\nu=-1$, and we use it to control the accuracy of the numerical integration.

\subsection{Effective potential}

From the normalization condition one can get
\begin{equation}
	f^{-1}\dot r^2+r^2\dot\theta^2=\frac{\mathcal E^2-V_{\rm eff}^2(r,\theta)}{f},
\end{equation}
with
\begin{equation}
	V_{\rm eff}(r,\theta)=\sqrt{f(r)\left[1+\frac{\big(\mathcal L-\beta\Psi(r)\sin^2\theta\big)^2}{r^2\sin^2\theta}\right]} .
	\label{eq:Veff}
\end{equation}
The particle can move only in the regions where $\mathcal E\ge V_{\rm eff}(r,\theta)$, and the boundaries of these regions are determined by $\mathcal E=V_{\rm eff}$. It is easy to see that $\partial_\theta V_{\rm eff}=0$ at $\theta=\pi/2$. Hence, the equatorial plane is a symmetry plane, and $\theta=\pi/2$, $\dot\theta=0$ is a consistent solution. In the equatorial plane we have $V_{\rm eff}^2=f(1+\mathcal X^2/r^2)$ with $\mathcal X=\mathcal L-\beta\Psi$.

\begin{figure*}[t]
	\centering
	\includegraphics[width=0.9\textwidth]{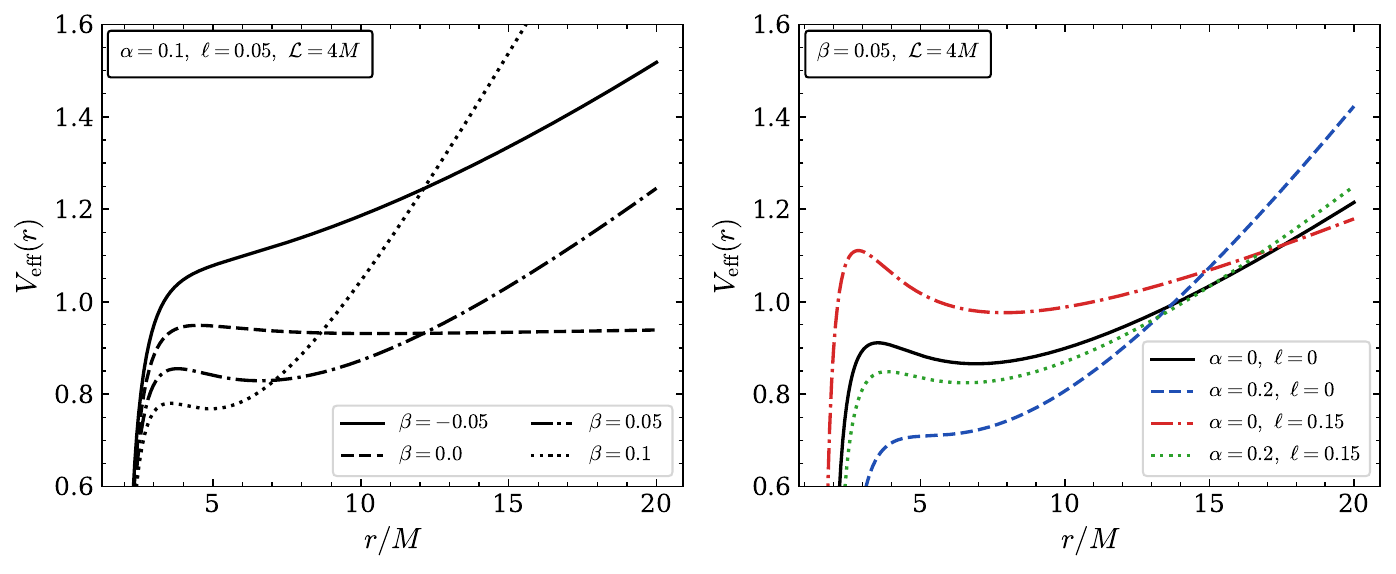}
	\caption{Equatorial effective potential $V_{\rm eff}(r)$ for $\mathcal L=4M$. Left: different values of the magnetic coupling $\beta$ for $\alpha=0.1$, $\ell=0.05$. Right: different $(\alpha,\ell)$ for $\beta=0.05$.}
	\label{fig:veff}
\end{figure*}

Fig.~\ref{fig:veff} shows the effective potential. Because $\Psi$ increases approximately as $r^2$, for any $\beta\neq0$ the term $\beta^2\Psi^2/r^2$ becomes dominant at large $r$. This term confines the motion in the equatorial plane, but the particle still can escape along the field lines ($\theta\to0,\pi$). For anti-Larmor coupling ($\beta>0$) the magnetic field creates a deep potential well, while for $\beta<0$ it raises the potential. From the right panel we see that at fixed $\beta$ and $\mathcal L$ the CoS ($\alpha>0$) and the KR parameter ($\ell>0$) shift the potential in opposite directions. For $A=1$ both the metric and the magnetic profile become those of Schwarzschild and Wald ($\Psi=r^2$). Thus, all configurations with $\alpha=\ell$ give exactly the same charged-particle dynamics as in the magnetized Schwarzschild case.

\subsection{Circular orbits and the ISCO}

\begin{figure*}[t]
	\centering
	\includegraphics[width=0.9\textwidth]{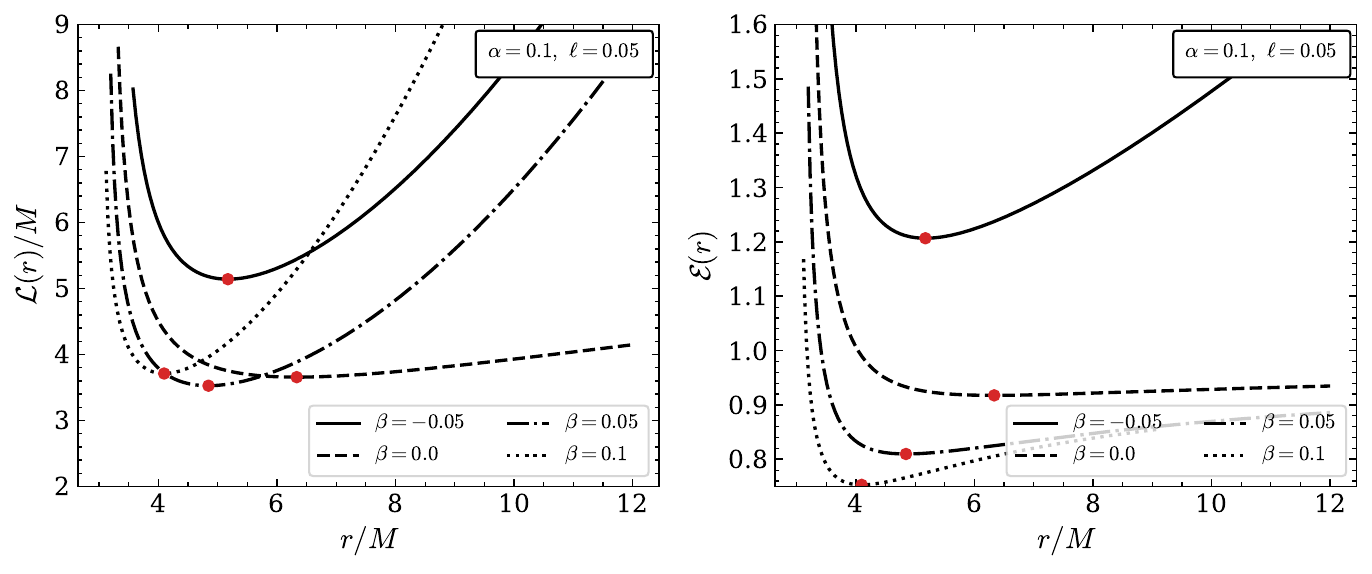}
	\caption{Specific canonical angular momentum $\mathcal L(r)$ (left) and specific energy $\mathcal E(r)$ (right) of charged particles on circular orbits for $\alpha=0.1$, $\ell=0.05$ and different values of $\beta$. Red dots mark the ISCO.}
	\label{fig:LE}
\end{figure*}

For circular orbits in the equatorial plane the conditions $V_{\rm eff}=\mathcal E$ and $\partial_rV_{\rm eff}=0$ must hold. The second condition leads to a quadratic equation for the mechanical angular momentum $\mathcal X=\mathcal L-\beta\Psi$,
\begin{equation}
	(rf'-2f)\,\mathcal X^2-2\beta f r\Psi'\,\mathcal X+r^3f'=0 ,
	\label{eq:Xquad}
\end{equation}
and its positive root, which corresponds to counter-clockwise motion, is
\begin{equation}
	\mathcal X(r)=\frac{\sqrt{\beta^2f^2r^2\Psi'^2+r^3f'(2f-rf')}-\beta f r\Psi'}{2f-rf'} .
	\label{eq:Xsol}
\end{equation}
Then the specific energy and the canonical angular momentum on the circular orbit are
\begin{equation}
	\mathcal E(r)=\sqrt{f\left(1+\frac{\mathcal X^2}{r^2}\right)},\qquad \mathcal L(r)=\mathcal X(r)+\beta\Psi(r).
	\label{eq:ELcirc}
\end{equation}
Circular orbits exist when $2f-rf'>0$, that is, outside the photon sphere $r_{\rm ph}=3M/A$. For neutral particles ($\beta=0$) we get $\mathcal X^2=Mr^2/(Ar-3M)$ and $\mathcal E^2=2f^2/(2f-rf')$. We plot $\mathcal L(r)$ and $\mathcal E(r)$ in Fig.~\ref{fig:LE}. The distance between the curves increases with $r$. The reason is that the canonical angular momentum includes the term $\beta\Psi\propto r^{s_+}$ coming from the vector potential, while the energy depends on $\beta$ only through $\mathcal X$.

\begin{figure*}[t]
	\centering
	\includegraphics[width=\textwidth]{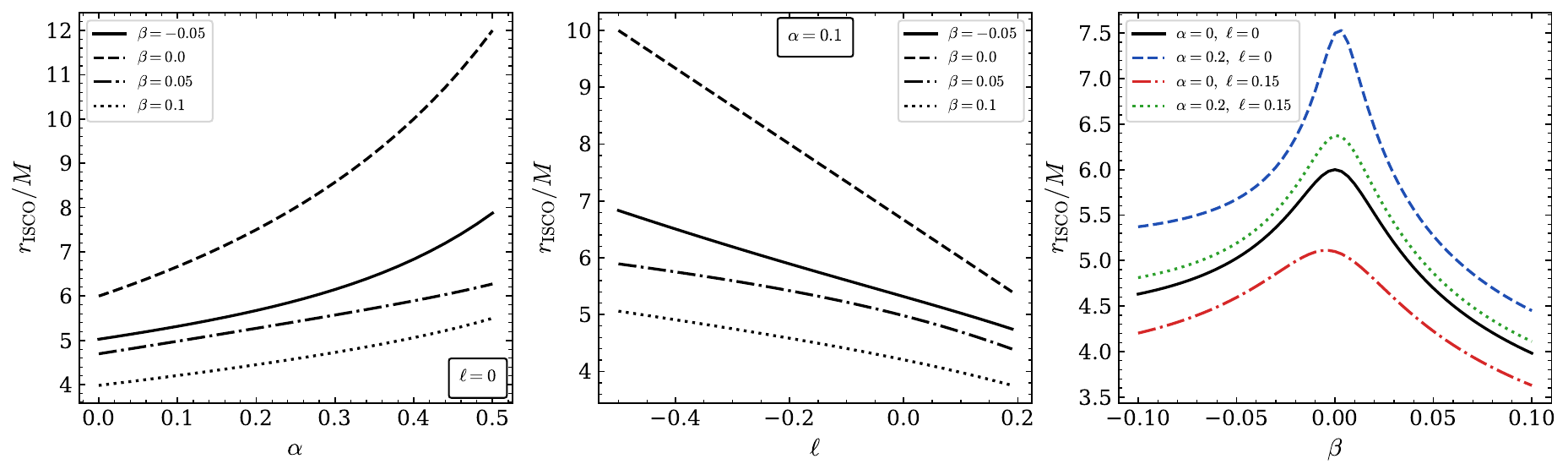}
	\caption{ISCO radius as a function of $\alpha$ for $\ell=0$ (left), of $\ell$ for $\alpha=0.1$ (middle), and of the magnetic coupling $\beta$ for different $(\alpha,\ell)$ (right).}
	\label{fig:isco}
\end{figure*}

Stable circular orbits correspond to the minima of $V_{\rm eff}$. The ISCO~\cite{Bardeen:1972fi} is the radius where $\partial^2_rV_{\rm eff}=0$ at fixed $\mathcal E$ and $\mathcal L$, or, in other words, where the radial epicyclic frequency (see Sec.~\ref{sec5}) becomes zero. For $\beta=0$ we find
\begin{equation}
	r_{\rm ISCO}=\frac{6M}{A}=\frac{6M(1-\ell)}{1-\alpha},\qquad \mathcal E_{\rm ISCO}=\sqrt{\frac{8A}{9}},
	\label{eq:isco0}
\end{equation}
It is seen that the ISCO moves outward due to the CoS and inward due to a positive KR parameter, but the ratio $r_{\rm ISCO}/r_{\rm h}=3$ stays the same. In the case $\beta\neq0$ we find the ISCO numerically, and the results are presented in Fig.~\ref{fig:isco} and in Table~\ref{tab:isco}. Similar to the Schwarzschild case~\cite{Frolov:2010mi,Kolos:2015iva}, the magnetic field shifts the ISCO inward for both signs of $\beta$, and the shift is larger for anti-Larmor coupling. As a test, for $A=1$ and $\beta=0$ our code gives $r_{\rm ISCO}=6M$, $\mathcal E=\sqrt{8/9}$, $\mathcal L=2\sqrt3M$ and $M\Omega=6^{-3/2}$ with machine precision.

\begin{table}[t]
\caption{ISCO radius, specific energy, specific canonical angular momentum and physical angular velocity $M\Omega_\phi^{\rm phys}=M\Omega_\phi/\sqrt A$ at the ISCO for selected values of $(\alpha,\ell)$ and of the magnetic coupling $\beta$ ($M=1$).}
\label{tab:isco}
\centering\footnotesize\setlength{\tabcolsep}{3.5pt}
\begin{adjustbox}{max width=\columnwidth}
\begin{tabular}{cccrcccc}
\toprule
$\alpha$ & $\ell$ & $A$ & $\beta$ & $r_{\rm ISCO}/M$ & $\mathcal E_{\rm ISCO}$ & $\mathcal L_{\rm ISCO}/M$ & $M\Omega^{\rm phys}_\phi$ \\
\midrule
0.00 & 0.00 & 1.0000 & $-0.05$ & 5.0257 & 1.1954 & 4.6268 & 0.11744 \\
0.00 & 0.00 & 1.0000 & $+0.00$ & 6.0000 & 0.9428 & 3.4641 & 0.06804 \\
0.00 & 0.00 & 1.0000 & $+0.05$ & 4.6967 & 0.8406 & 3.3592 & 0.06986 \\
0.00 & 0.00 & 1.0000 & $+0.10$ & 3.9827 & 0.7845 & 3.5223 & 0.07746 \\
\midrule
0.10 & 0.00 & 0.9000 & $-0.05$ & 5.3172 & 1.2249 & 5.7199 & 0.11576 \\
0.10 & 0.00 & 0.9000 & $+0.00$ & 6.6667 & 0.8944 & 3.8490 & 0.06124 \\
0.10 & 0.00 & 0.9000 & $+0.05$ & 4.9808 & 0.7809 & 3.6946 & 0.06384 \\
0.10 & 0.00 & 0.9000 & $+0.10$ & 4.2056 & 0.7234 & 3.9027 & 0.07098 \\
\midrule
0.20 & 0.00 & 0.8000 & $-0.05$ & 5.6746 & 1.3028 & 7.5248 & 0.11310 \\
0.20 & 0.00 & 0.8000 & $+0.00$ & 7.5000 & 0.8433 & 4.3301 & 0.05443 \\
0.20 & 0.00 & 0.8000 & $+0.05$ & 5.2723 & 0.7155 & 4.1147 & 0.05808 \\
0.20 & 0.00 & 0.8000 & $+0.10$ & 4.4504 & 0.6569 & 4.3930 & 0.06442 \\
\midrule
0.00 & 0.10 & 1.1111 & $-0.05$ & 4.7459 & 1.1915 & 3.8367 & 0.11904 \\
0.00 & 0.10 & 1.1111 & $+0.00$ & 5.4000 & 0.9938 & 3.1177 & 0.07560 \\
0.00 & 0.10 & 1.1111 & $+0.05$ & 4.3939 & 0.9018 & 3.0549 & 0.07689 \\
0.00 & 0.10 & 1.1111 & $+0.10$ & 3.7544 & 0.8472 & 3.1860 & 0.08468 \\
\midrule
0.10 & 0.05 & 0.9474 & $-0.05$ & 5.1730 & 1.2067 & 5.1409 & 0.11662 \\
0.10 & 0.05 & 0.9474 & $+0.00$ & 6.3333 & 0.9177 & 3.6566 & 0.06446 \\
0.10 & 0.05 & 0.9474 & $+0.05$ & 4.8451 & 0.8098 & 3.5271 & 0.06665 \\
0.10 & 0.05 & 0.9474 & $+0.10$ & 4.0977 & 0.7530 & 3.7115 & 0.07405 \\
\midrule
0.20 & 0.15 & 0.9412 & $-0.05$ & 5.1911 & 1.2086 & 5.2095 & 0.11652 \\
0.20 & 0.15 & 0.9412 & $+0.00$ & 6.3750 & 0.9147 & 3.6806 & 0.06404 \\
0.20 & 0.15 & 0.9412 & $+0.05$ & 4.8627 & 0.8061 & 3.5480 & 0.06628 \\
0.20 & 0.15 & 0.9412 & $+0.10$ & 4.1116 & 0.7492 & 3.7352 & 0.07365 \\
\bottomrule
\end{tabular}
\end{adjustbox}
\end{table}

\section{Trajectories of charged particles}\label{sec4}

In this section we study the trajectories of charged particles and how they are affected by the CoS, the Lorentz-violating KR field and the magnetic field. To obtain the trajectories we integrate the Hamilton equations following from Eq.~\eqref{eq:Ham} using the adaptive explicit Runge--Kutta method of order 8(5,3) (DOP853), with relative and absolute tolerances $10^{-10}$ and $10^{-12}$. For all trajectories presented here the constraint $|H|$ is less than $10^{-7}$ (usually $\lesssim10^{-11}$). We stop the integration if the particle reaches $r=1.01\,r_{\rm h}$ (capture) or $|z|=40M$ (escape along the field lines). Three cases are compared: the magnetized Schwarzschild black hole ($\alpha=\ell=0$, $A=1$), a case where the CoS dominates ($\alpha=0.15$, $\ell=0$, $A=0.85$) and a case where the KR field dominates ($\alpha=0$, $\ell=0.1$, $A=1.11$). The allowed regions are bounded by the curves $\mathcal E=V_{\rm eff}(r,\theta)$, Eq.~\eqref{eq:Veff}.

\begin{figure*}[t]
	\centering
	\includegraphics[width=\textwidth]{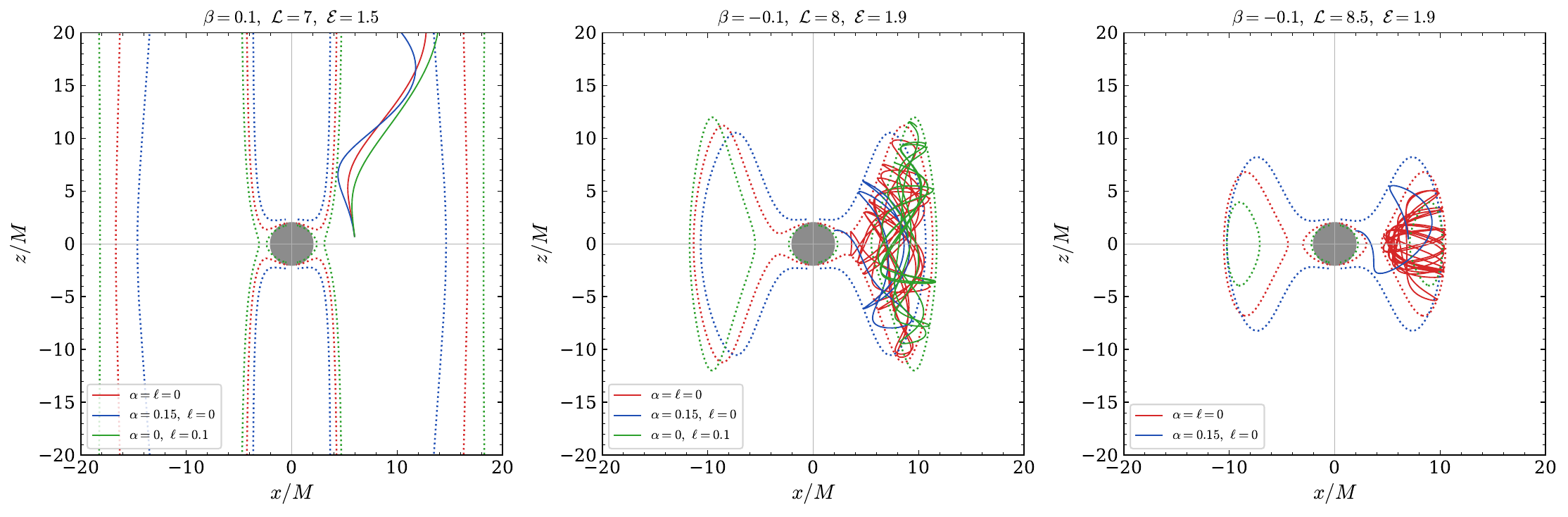}
	\caption{Trajectories of charged particles projected onto the $(x,z)$ plane together with the boundaries $\mathcal E=V_{\rm eff}$ (dotted) for $(\alpha,\ell)=(0,0)$ (red), $(0.15,0)$ (blue) and $(0,0.1)$ (green). Left: escape along the magnetic field lines. Middle and right: trapped chaotic motion and capture by the black hole. The grey disk marks the region $r<2M$.}
	\label{fig:regimes}
\end{figure*}

In Fig.~\ref{fig:regimes} we show three typical kinds of motion in the $(x,z)$ plane, where $x=r\sin\theta$ and $z=r\cos\theta$. The left panel corresponds to anti-Larmor coupling with high energy ($\beta=0.1$, $\mathcal L=7$, $\mathcal E=1.5$, initial position $r_0=6M$, $\theta_0=1.45$). In this case the allowed region is open along the field lines, and in all three spacetimes the particle escapes to $|z|\to\infty$; the parameter $A$ changes only the shape of the escape channel. The middle and right panels correspond to Larmor coupling ($\beta=-0.1$, $\mathcal E=1.9$, $\mathcal L=8$ and $8.5$). Here the fate of the particle depends on the geometry. With the same initial conditions, for $A=1$ and $A=1.11$ the particle moves on a trapped chaotic orbit inside a toroidal region around the black hole, but for $A=0.85$ (CoS-dominated case) the inner and outer allowed regions join and the particle falls into the black hole. For $\mathcal L=8.5$ and $A=1.11$ the initial point is in a forbidden region, so there is no such trajectory. We conclude that small changes of $\alpha$ and $\ell$ can change the topology of the allowed region. A similar behaviour was found for the KR black hole in Ref.~\cite{Rodrigues:2026ofx}, while in our case it is controlled by only one parameter $A$.

\begin{figure*}[t]
	\centering
	\includegraphics[width=0.92\textwidth]{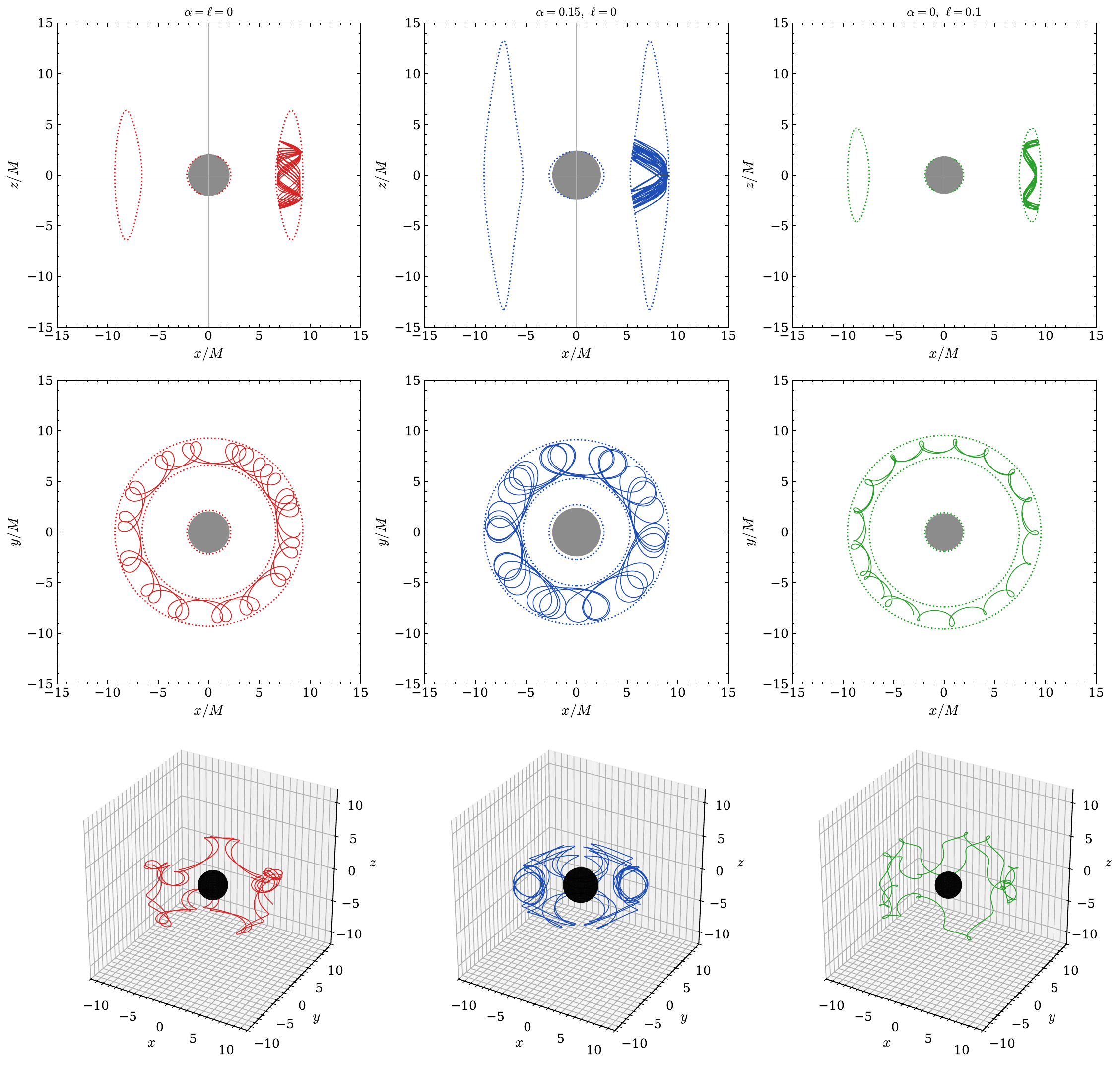}
	\caption{Curled trajectories of charged particles for $\beta=0.1$, $\mathcal L=7$, $r_0=9M$ and $\theta_0=1.6$, in the $(x,z)$ plane (top row, with the boundaries $\mathcal E=V_{\rm eff}$ dotted), in the equatorial $(x,y)$ plane (middle row) and in three dimensions (bottom row). Columns: $(\alpha,\ell)=(0,0)$, $(0.15,0)$ and $(0,0.1)$. The energy is $\mathcal E=1.012\,V_{\rm eff}(r_0,\pi/2)$, i.e.\ $\mathcal E=0.8991$, $0.8471$ and $0.9542$, respectively.}
	\label{fig:curled}
\end{figure*}

Among bounded motions, \emph{curled} (epicyclic) trajectories are of special interest. For them the azimuthal velocity $\dot\phi$ changes its sign along the orbit. From Eq.~\eqref{eq:tphidot} it follows that in the equatorial plane this happens at the radius $r_*$ given by $\mathcal L=\beta\Psi(r_*)$. This is possible only if $\beta\mathcal L>0$, i.e., for anti-Larmor motion~\cite{Kolos:2015iva,Rodrigues:2026ofx}. When $A=1$ we have simply $r_*=\sqrt{\mathcal L/\beta}$. Curled trajectories for $\beta=0.1$, $\mathcal L=7$, $r_0=9M$, $\theta_0=1.6$ and energy $1.2\%$ higher than the equatorial potential at $r_0$ are shown in Fig.~\ref{fig:curled}. The particle moves inside a thin toroidal region, and in the projection onto the equatorial plane one can see the typical loops. The number and size of the loops depend on $A$, both through the gravitational potential and through the magnetic profile $\Psi$. In the CoS-dominated case ($A=0.85$) the allowed region is wider, the vertical oscillations are larger and the loops are broader. In the KR-dominated case ($A=1.11$) the region is narrower and the curls are tighter and more regular. Such quasi-periodic motion is the reason why we study harmonic oscillations in the next section.

\section{Quasi-periodic oscillations}\label{sec5}

\subsection{Fundamental frequencies}

Let us consider a charged particle moving on a stable circular orbit with radius $r_0$ in the equatorial plane, and let it be slightly perturbed, $r=r_0+\delta r$, $\theta=\pi/2+\delta\theta$, with $\mathcal E$ and $\mathcal L$ kept fixed. Expanding the Hamiltonian~\eqref{eq:Ham} up to second order~\cite{Kolos:2015iva,Tursunov:2016dss}, we obtain the harmonic oscillator equations $\delta\ddot r+\omega_r^2\delta r=0$ and $\delta\ddot\theta+\omega_\theta^2\delta\theta=0$ (dots mean $d/d\tau$), where $\omega_r^2=\frac12\partial_r^2V_{\rm eff}^2$ and $\omega_\theta^2=r^{-2}\partial_\theta^2\big[\mathcal X^2/(2r^2\sin^2\theta)\big]$. After dividing by $\dot t^2=\mathcal E^2/f^2$, the frequencies with respect to the coordinate time are
\begin{align}
	\Omega_\phi&=\frac{d\phi}{dt}=\frac{f\,\mathcal X}{\mathcal E\,r^2}, \label{eq:Omphi}\\
	\Omega_r^2&=\frac{f^2}{2\mathcal E^2}\,\frac{d^2}{dr^2}\Big[f\Big(1+\frac{\mathcal X^2}{r^2}\Big)\Big]_{\mathcal L}, \label{eq:Omr}\\
	\Omega_\theta^2&=\frac{f^2}{\mathcal E^2}\,\frac{\mathcal L^2-\beta^2\Psi^2}{r^4}, \label{eq:Omth}
\end{align}
Here all quantities are taken on the circular orbit, Eqs.~\eqref{eq:Xsol}--\eqref{eq:ELcirc}. The derivative in Eq.~\eqref{eq:Omr} is calculated at fixed $\mathcal L$, and therefore $\mathcal X'=-\beta\Psi'$ and $\mathcal X''=-\beta\Psi''$. In explicit form,
\begin{align}
	\frac{d^2V_{\rm eff}^2}{dr^2}={}&f''\Big(1+\frac{\mathcal X^2}{r^2}\Big)+2f'\Big(\frac{2\mathcal X\mathcal X'}{r^2}-\frac{2\mathcal X^2}{r^3}\Big)\nonumber\\
	&+f\Big(\frac{2\mathcal X'^2+2\mathcal X\mathcal X''}{r^2}-\frac{8\mathcal X\mathcal X'}{r^3}+\frac{6\mathcal X^2}{r^4}\Big).
	\label{eq:Vpp}
\end{align}
In the case $\beta=0$ these expressions become simply
\begin{equation}
	\Omega_\phi^2=\Omega_\theta^2=\frac{M}{r^3},\qquad \Omega_r^2=\frac{M}{r^3}\Big(A-\frac{6M}{r}\Big),
	\label{eq:Om0}
\end{equation}
and $\Omega_r$ is zero at $r_{\rm ISCO}=6M/A$, which agrees with Eq.~\eqref{eq:isco0}. When the magnetic field is present, $\Omega_\theta\neq\Omega_\phi$, since the vertical restoring force includes the Lorentz term $\propto\beta^2\Psi^2$ in Eq.~\eqref{eq:Omth}.

Using Eq.~\eqref{eq:tphys}, the frequencies measured by a distant observer are $\Omega_i^{\rm phys}=\Omega_i/\sqrt A$. In physical units they are
\begin{align}
	\nu_i&=\frac{1}{2\pi}\frac{c^3}{GM}\frac{\bar\Omega_i}{\sqrt A}\nonumber\\
	&\simeq 3.231\times10^{4}\,{\rm Hz}\,\Big(\frac{M_\odot}{M}\Big)\frac{\bar\Omega_i}{\sqrt A},\quad i=\phi,r,\theta ,
	\label{eq:nuHz}
\end{align}
where $\bar\Omega_i=M\Omega_i$ is a dimensionless quantity. For neutral particles, from Eqs.~\eqref{eq:Om0} and \eqref{eq:nuHz} one gets $\nu_\phi\propto(M/A)^{-1}$ at fixed $Ar/M$. This means that far from the black hole the Keplerian motion looks like motion around a body with effective mass $M/A$, while the ratio $\nu_r/\nu_\phi=\sqrt{A-6M/r}$ depends on $A$ in a different way.

We checked Eqs.~\eqref{eq:Omphi}--\eqref{eq:Omth} in three different ways. First, symbolically, by obtaining Eq.~\eqref{eq:Om0} in the limit $\beta=0$. Second, we compared the analytical $\partial_r^2V_{\rm eff}^2$ with its finite-difference value (the relative difference is $<10^{-6}$). Third, we integrated the full nonlinear equations of motion for slightly perturbed circular orbits (for example, $A=0.95$, $\beta=0.03$, $r_0=8M$ and $A=1.05$, $\beta=-0.02$, $r_0=9M$) and found the frequencies from the Fourier spectra of $r(t)$ and $\theta(t)$. The numerical and analytical values of $\Omega_\phi$, $\Omega_r$ and $\Omega_\theta$ coincide within the spectral resolution, that is, with relative accuracy better than $3\times10^{-4}$.

\begin{figure*}[t]
	\centering
	\includegraphics[width=\textwidth]{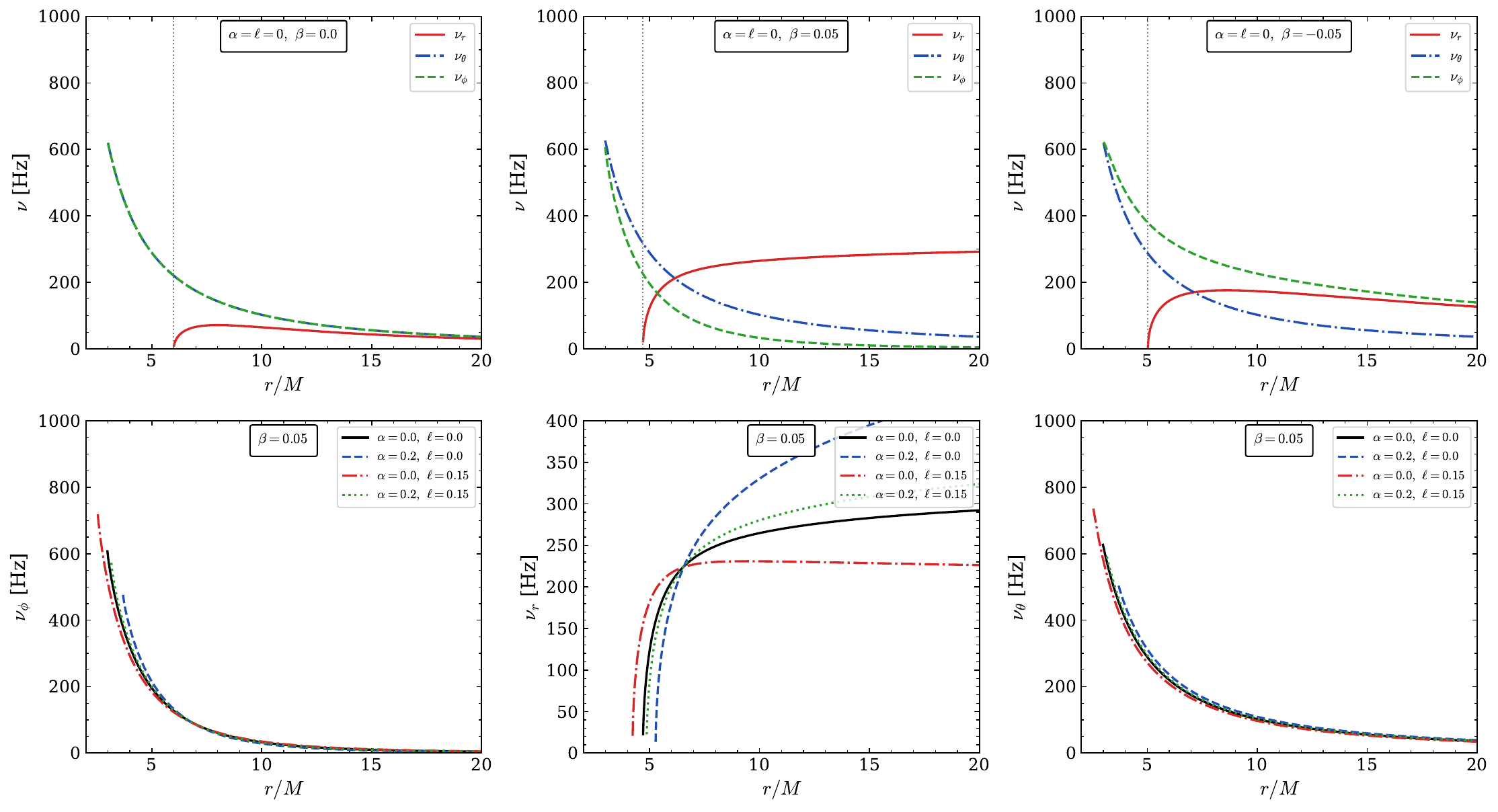}
	\caption{Fundamental frequencies $\nu_\phi$, $\nu_r$ and $\nu_\theta$ (in Hz, for $M=10M_\odot$). Top row: Schwarzschild geometry ($\alpha=\ell=0$) for $\beta=0$, $0.05$ and $-0.05$; the vertical dotted line marks the ISCO. Bottom row: $\nu_\phi$, $\nu_r$ and $\nu_\theta$ for $\beta=0.05$ and different $(\alpha,\ell)$.}
	\label{fig:freqs}
\end{figure*}

Fig.~\ref{fig:freqs} shows the three frequencies for $M=10M_\odot$. For $\beta=0$ and $A=1$ the orbital and vertical frequencies are equal, and the radial frequency is zero at $r=6M$. The magnetic field removes this degeneracy between $\nu_\phi$ and $\nu_\theta$ and changes $\nu_r$ considerably. For anti-Larmor coupling ($\beta>0$) the Lorentz force partly compensates gravity, so the orbital frequency becomes smaller and the radial frequency can be larger than it. For Larmor coupling ($\beta<0$) the situation is opposite. At fixed $\beta$, the CoS (smaller $A$) moves the radius where radial stability begins outward, and the KR parameter moves it inward. At the same time, $\nu_\phi$ and $\nu_\theta$ change only slightly at fixed $r/M$.

\subsection{Relativistic precession model and MCMC analysis}

In the RPM~\cite{Stella:1997tc,Stella:1999sj}, the upper and lower high-frequency QPOs of a twin-peak pair are associated with the orbital frequency and the periastron-precession frequency at the emission radius $r$,
\begin{equation}
	\nu_{\rm U}^{\rm th}=\nu_\phi ,\qquad \nu_{\rm L}^{\rm th}=\nu_\phi-\nu_r .
	\label{eq:rpm}
\end{equation}
In our analysis we use the twin-peak QPOs of three microquasars given in Table~\ref{tab:data}: GRO J1655$-$40, with $\nu_{\rm U}=450\pm3$~Hz and $\nu_{\rm L}=300\pm5$~Hz~\cite{Strohmayer:2001yn,Remillard:2002cy}; XTE J1550$-$564, with $276\pm3$~Hz and $184\pm5$~Hz~\cite{Remillard:2002cy}; and GRS 1915$+$105, with $168\pm3$~Hz and $113\pm5$~Hz~\cite{Remillard:2006fc,Belloni:2012sv}. For the black-hole masses we use the values from dynamical measurements: $M=5.4\pm0.3M_\odot$ for GRO J1655$-$40~\cite{Beer:2001cg} (see also Refs.~\cite{Greene:2001wd,Motta:2013wga}), $M=9.1\pm0.6M_\odot$ for XTE J1550$-$564~\cite{Orosz:2011ki} and $M=12.4^{+2.0}_{-1.8}M_\odot$ for GRS 1915$+$105~\cite{Reid:2014ywa}.

\begin{table}[t]
\caption{Twin-peak QPO frequencies and dynamical masses of the three microquasars used in the MCMC analysis.}
\label{tab:data}
\centering\small
\begin{adjustbox}{max width=\columnwidth}
\begin{tabular}{lccc}
\toprule
Source & $\nu_{\rm U}$ [Hz] & $\nu_{\rm L}$ [Hz] & $M/M_\odot$ \\
\midrule
GRO J1655$-$40\cite{Strohmayer:2001yn,Remillard:2002cy,Beer:2001cg} & $450\pm3$ & $300\pm5$ & $5.4\pm0.3$ \\
XTE J1550$-$564\cite{Remillard:2002cy,Orosz:2011ki} & $276\pm3$ & $184\pm5$ & $9.10\pm0.61$ \\
GRS 1915$+$105\cite{Remillard:2006fc,Belloni:2012sv,Reid:2014ywa} & $168\pm3$ & $113\pm5$ & $12.4^{+2.0}_{-1.8}$ \\
\bottomrule
\end{tabular}
\end{adjustbox}
\end{table}

According to the Bayes theorem, the posterior distribution of the parameter vector ${\boldsymbol\theta}$ is
\begin{equation}
	\mathcal P(\boldsymbol\theta|\mathcal D)\propto\mathcal L(\mathcal D|\boldsymbol\theta)\,\pi(\boldsymbol\theta),
	\label{eq:bayes}
\end{equation}
where for the two QPO frequencies we take the Gaussian log-likelihood
\begin{equation}
	\ln\mathcal L=-\sum_{i={\rm U,L}}\left[\frac12\left(\frac{\nu_i^{\rm obs}-\nu_i^{\rm th}(\boldsymbol\theta)}{\sigma_i}\right)^2+\ln\big(\sigma_i\sqrt{2\pi}\big)\right],
	\label{eq:lnL}
\end{equation}
and the priors are truncated Gaussians,
\begin{equation}
	\ln\pi(\boldsymbol\theta)=-\frac12\sum_j\left(\frac{\theta_j-\mu_j}{s_j}\right)^2\quad\text{for }\theta_j\in[\theta_j^{\rm low},\theta_j^{\rm high}],
	\label{eq:prior}
\end{equation}
with $\ln\pi=-\infty$ outside this box. The values of the prior parameters are given in Table~\ref{tab:priors}. For the mass we use the dynamical mass of each source as a prior. The CoS parameter is taken in the physical range $\alpha\ge0$, and the KR parameter in the interval $-0.70\le\ell\le0.19$ obtained from observations in Ref.~\cite{Junior:2024ety}; for both we use Gaussian widths $0.1$ with centres at $\alpha=0.05$ and $\ell=0$. The prior for the magnetic coupling is centred at $\beta=0$, and for the emission radius at $r=6M$. We also require that the orbit is outside the horizon, is stable ($\Omega_r^2>0$) and gives $\nu_{\rm L}>0$. It should be noted that the likelihood depends on $(\alpha,\ell)$ only through $A$. Therefore, the posteriors of $\alpha$ and $\ell$ separately are determined by the priors along the degeneracy line $A={\rm const}$, and the quantity which is really constrained by the data is $A$.

\begin{table}[t]
\caption{Truncated Gaussian priors $\mathcal N(\mu,s)$ on $[\theta^{\rm low},\theta^{\rm high}]$ adopted in the MCMC analysis.}
\label{tab:priors}
\centering\small
\begin{adjustbox}{max width=\columnwidth}
\begin{tabular}{lcc}
\toprule
Parameter & $(\mu,\ s)$ & $[\theta^{\rm low},\theta^{\rm high}]$ \\
\midrule
$M/M_\odot$ (GRO J1655$-$40) & $(5.4,\ 0.3)$ & $[4.8,\ 6.0]$ \\
$M/M_\odot$ (XTE J1550$-$564) & $(9.1,\ 0.3)$ & $[8.5,\ 9.7]$ \\
$M/M_\odot$ (GRS 1915$+$105) & $(12.4,\ 1.9)$ & $[8.6,\ 16.2]$ \\
$\alpha$ & $(0.05,\ 0.1)$ & $[0.0,\ 0.5]$ \\
$\ell$ & $(0.0,\ 0.1)$ & $[-0.7,\ 0.19]$ \\
$\beta$ & $(0.0,\ 0.05)$ & $[-0.1,\ 0.1]$ \\
$r/M$ & $(6.0,\ 1.5)$ & $[3.0,\ 10.0]$ \\
\bottomrule
\end{tabular}
\end{adjustbox}
\end{table}

We consider four nested models: (a) Schwarzschild, $\boldsymbol\theta=(M,r)$; (b) Letelier--KR without magnetic field, $\boldsymbol\theta=(M,\alpha,\ell,r)$; (c) magnetized Schwarzschild, $\boldsymbol\theta=(M,\beta,r)$; and (d) the full model, $\boldsymbol\theta=(M,\alpha,\ell,\beta,r)$. For sampling the posteriors we use the affine-invariant ensemble sampler~\cite{Goodman:2010et} of the \texttt{emcee} package~\cite{Foreman-Mackey:2012any} with $50$ walkers and $6\times10^4$ steps in each run ($10^5$ steps for the magnetized models of GRS 1915$+$105). The walkers start in a small ball around the maximum a posteriori point, which we find by a multi-start Nelder--Mead search. The first $1/6$ of each chain is removed as burn-in, and the rest of the chains are thinned by half of the shortest integrated autocorrelation time $\tau_{\rm int}$. The corner plots are made with \texttt{corner.py}~\cite{Foreman-Mackey:2016corner}.

One point here needs special care. The posteriors of the magnetized models (c) and (d) are \emph{bimodal}. Besides the main Larmor solution ($\beta<0$), the QPO pairs can also be fitted by an anti-Larmor solution ($\beta>0$) at somewhat smaller radii. Ensemble samplers cannot move walkers between modes that are far from each other, so a single \texttt{emcee} run does not give the relative weight of the modes. For this reason, we run \texttt{emcee} separately in the two regions $\beta<0$ and $\beta>0$. Then we find the posterior mass $w_+$ of the $\beta>0$ mode using the nested-sampling algorithm~\cite{Skilling:2006gxv} of the \texttt{dynesty} package~\cite{Speagle:2019ivv} (800 live points, multi-ellipsoidal bounds, random-walk proposals, stopping criterion $\Delta\ln Z<0.01$). This algorithm covers the whole prior volume and works well for multimodal distributions. The final posterior is the combination of the two \texttt{emcee} samples with weights $1-w_+$ and $w_+$. In addition, nested sampling gives the Bayesian evidence $Z=\int\mathcal L\,\pi\,d\boldsymbol\theta$ of each model, and from it we calculate the Bayes factors $\ln B_{ij}=\ln Z_i-\ln Z_j$. We also checked that the marginalized quantiles of the combined \texttt{emcee} posteriors are in agreement with the equally weighted \texttt{dynesty} posterior samples (see Sec.~\ref{sec:results}).

For the model comparison, we include the dynamical mass measurement as one more Gaussian data point, so that for each source we have $N=3$ data points. Then we compute the Akaike and Bayesian information criteria~\cite{Akaike:1974ic,Schwarz:1978bic,Liddle:2007fy}
\begin{equation}
	{\rm AIC}=2k-2\ln\mathcal L_{\max},\qquad {\rm BIC}=k\ln N-2\ln\mathcal L_{\max},
	\label{eq:aic}
\end{equation}
where $k$ is the number of free parameters and $\mathcal L_{\max}$ is the maximum value of the extended likelihood over the posterior samples. The values $\Delta{\rm AIC}$ and $\Delta{\rm BIC}$ are given with respect to the best model for each source. Usually, $\Delta\lesssim2$ means that the models are statistically equivalent, $4\lesssim\Delta\lesssim7$ means considerably less support, and $\Delta\gtrsim10$ means almost no support~\cite{Liddle:2007fy}. Since $N$ is very small, these criteria can be used only as an indication. The Bayes factors automatically penalize the prior volume of extra parameters, and they give a more reliable, complementary measure. We use the empirical Jeffreys scale of Ref.~\cite{Trotta:2008qt}: $|\ln B|<1$ is inconclusive, $1<|\ln B|<2.5$ is weak, $2.5<|\ln B|<5$ is moderate and $|\ln B|>5$ is strong evidence.

\subsection{Results}\label{sec:results}

\begin{figure*}[t]
	\centering
	\includegraphics[width=0.78\textwidth]{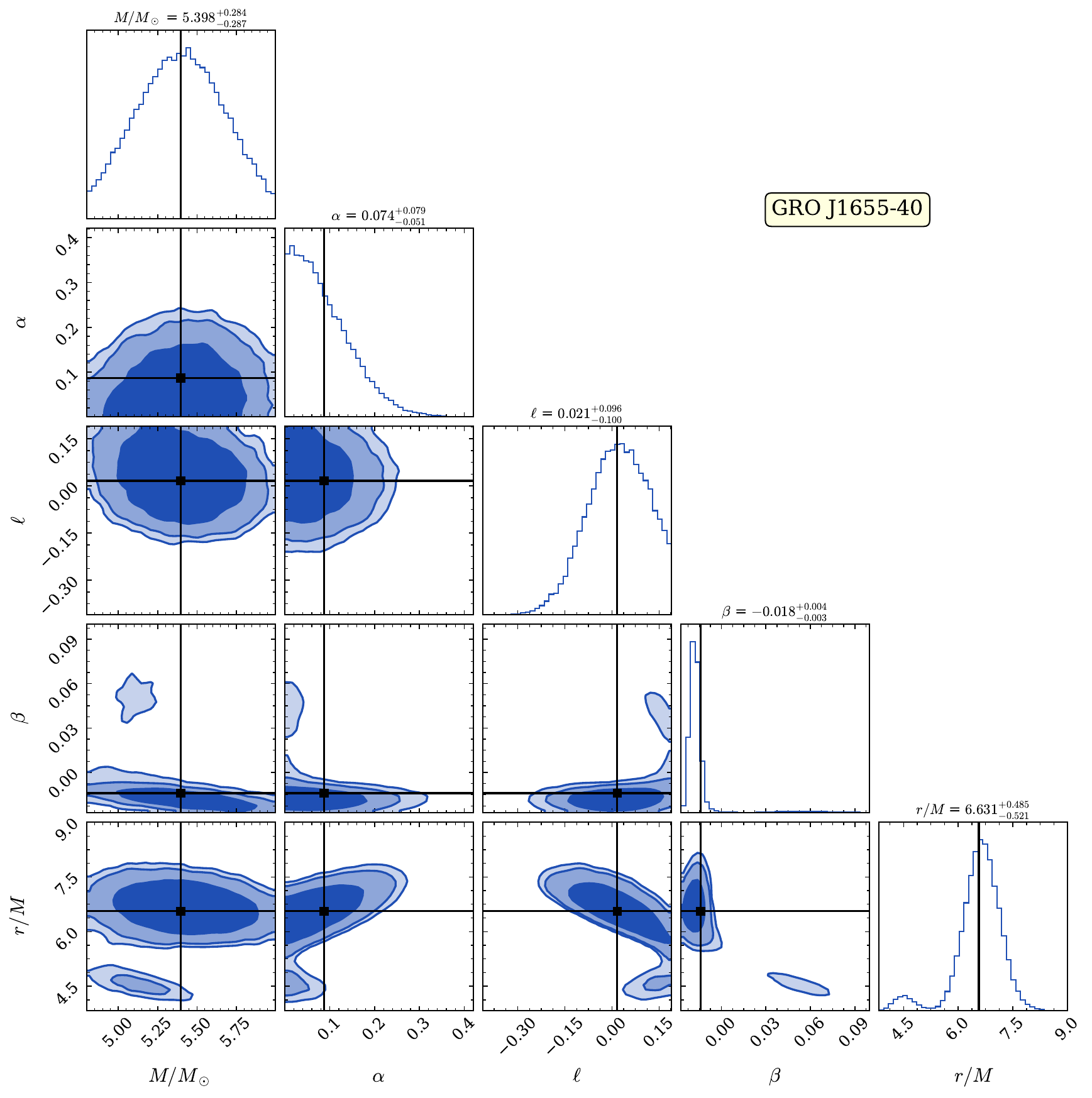}
	\caption{One- and two-dimensional marginalized posterior distributions of $M/M_\odot$, $\alpha$, $\ell$, $\beta$ and $r/M$ for GRO J1655$-$40 (full model). Contours enclose $68\%$, $90\%$, and $95\%$ of the probability; the black lines mark the posterior means.}
	\label{fig:cornerGRO}
\end{figure*}

\begin{figure*}[t]
	\centering
	\includegraphics[width=0.78\textwidth]{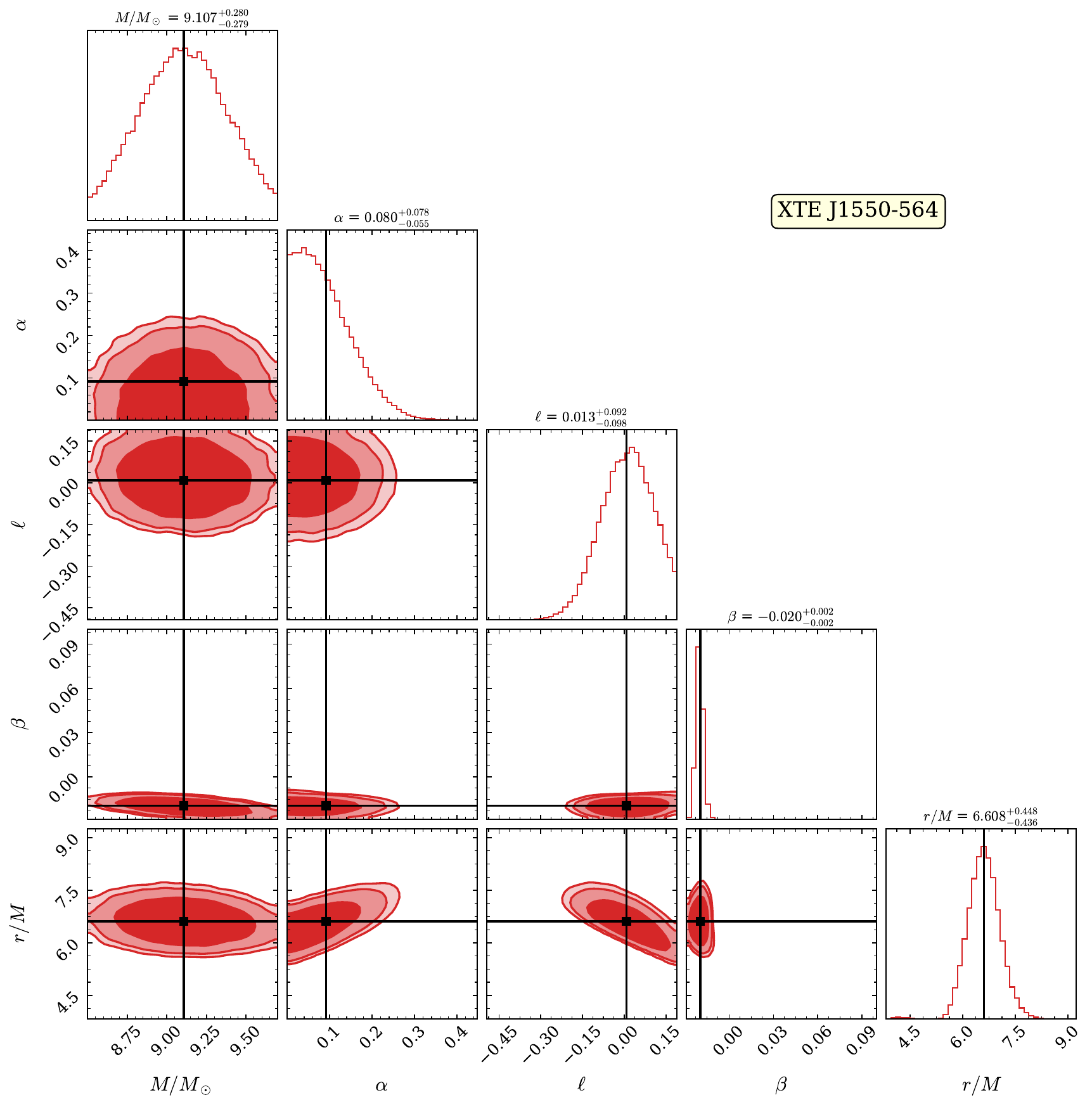}
	\caption{Same as Fig.~\ref{fig:cornerGRO} for XTE J1550$-$564.}
	\label{fig:cornerXTE}
\end{figure*}

\begin{figure*}[t]
	\centering
	\includegraphics[width=0.78\textwidth]{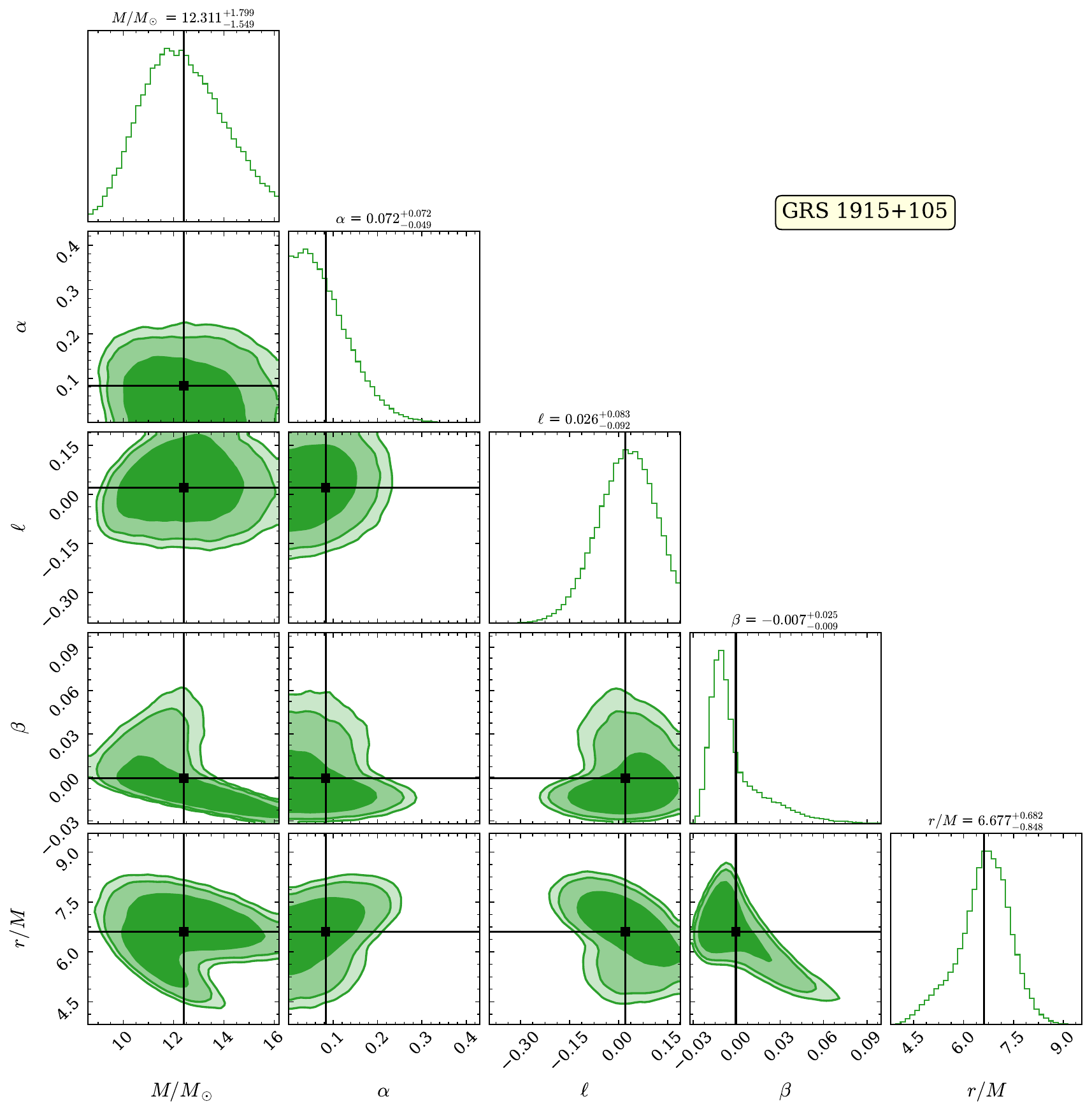}
	\caption{Same as Fig.~\ref{fig:cornerGRO} for GRS 1915$+$105. The secondary anti-Larmor ($\beta>0$) mode carries $33\%$ of the posterior mass.}
	\label{fig:cornerGRS}
\end{figure*}

\begin{figure*}[t]
	\centering
	\includegraphics[width=\textwidth]{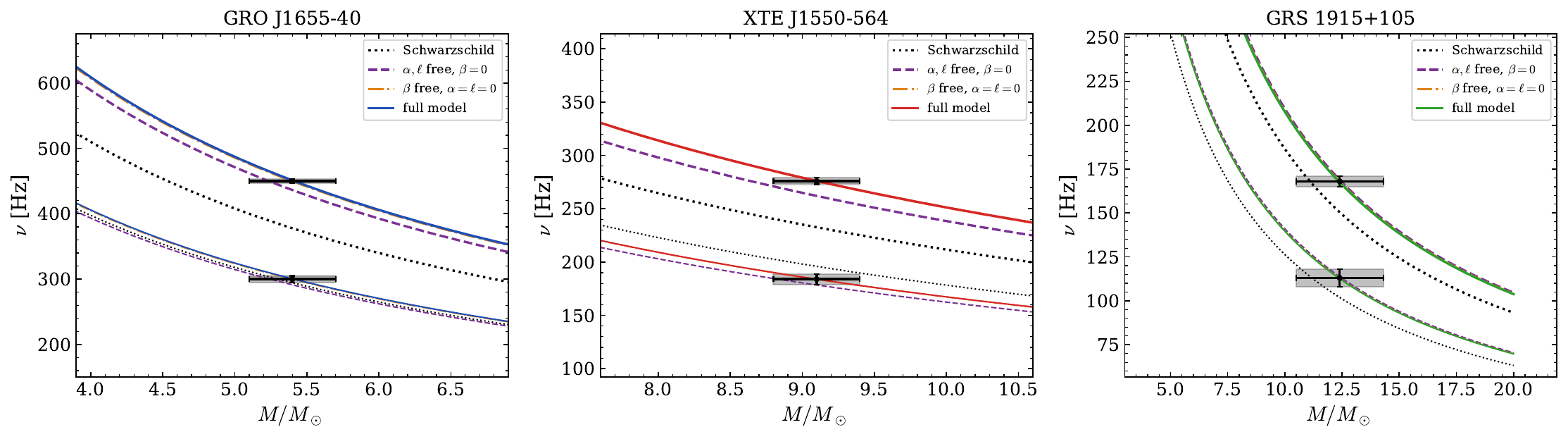}
	\caption{Upper ($\nu_{\rm U}=\nu_\phi$, upper curves) and lower ($\nu_{\rm L}=\nu_\phi-\nu_r$, lower curves) RPM frequencies as functions of the black-hole mass, evaluated at the maximum-likelihood values of the dimensionless parameters of the four models. The grey boxes show the observed frequencies ($\pm1\sigma$) and the dynamical masses ($\pm1\sigma$ of the prior).}
	\label{fig:qpomass}
\end{figure*}

\begin{figure*}[t]
	\centering
	\includegraphics[width=2\columnwidth]{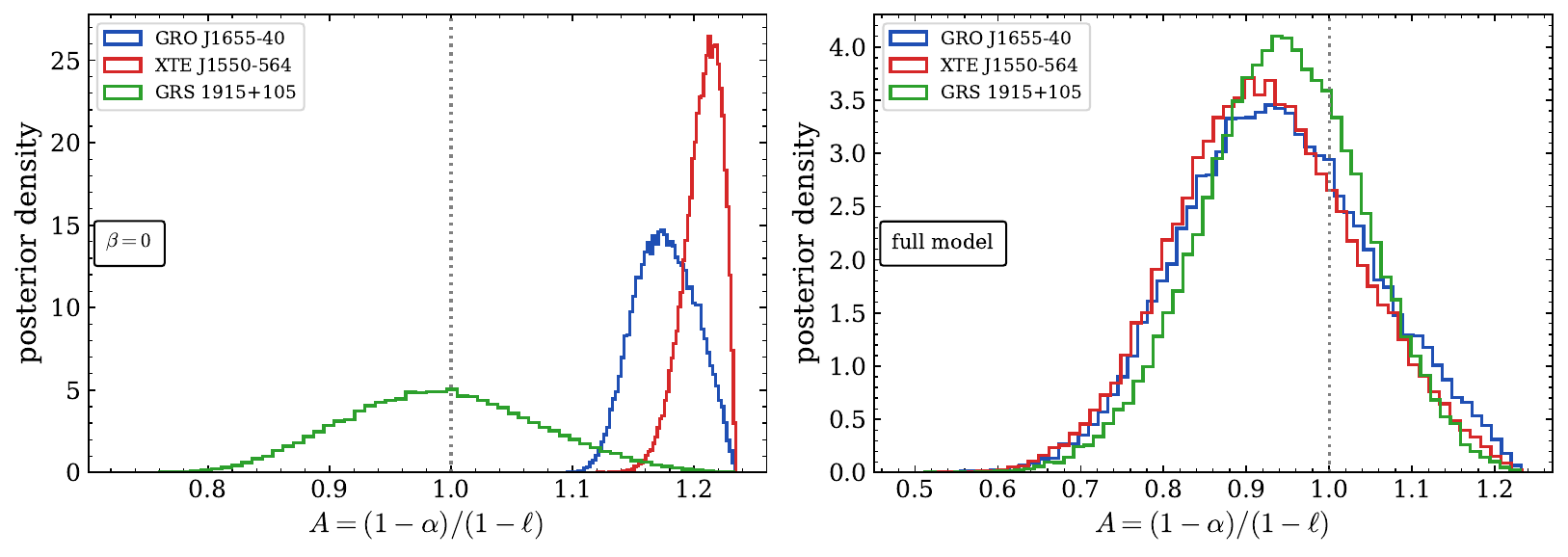}
	\caption{Posterior distributions of $A=(1-\alpha)/(1-\ell)$ without (left) and with (right) the magnetic field. The dotted line marks the Schwarzschild value $A=1$.}
	\label{fig:Apost}
\end{figure*}

Table~\ref{tab:results} contains the marginalized constraints for all models and sources, Table~\ref{tab:evidence} gives the Bayesian evidences, and Table~\ref{tab:conv} gives the convergence diagnostics of the full model. In Figs.~\ref{fig:cornerGRO}--\ref{fig:cornerGRS} we show the one- and two-dimensional posteriors of the full model; the contours correspond to $68\%$, $90\%$ and $95\%$ of the probability, and the black lines show the posterior means. In Fig.~\ref{fig:qpomass} the best-fit RPM frequencies of the four models are compared with the data as functions of the black-hole mass, and Fig.~\ref{fig:Apost} presents the posterior of the derived parameter $A$.

\emph{Convergence and robustness.} In all runs the number of steps is at least $77$ times larger than the largest integrated autocorrelation time, and in most runs it is several hundred times larger (Table~\ref{tab:conv}). The mean acceptance fractions are between $0.23$ and $0.71$. For all parameters, models and sources, the posterior medians and $68\%$ intervals of the combined \texttt{emcee} samples agree with the results of the independent \texttt{dynesty} runs, usually within a few per cent of the interval width. For example, for the full model of XTE J1550$-$564 we get $\beta=-0.0199^{+0.0024}_{-0.0023}$ from \texttt{emcee} and $-0.0197^{+0.0021}_{-0.0024}$ from \texttt{dynesty}. The best-fit frequencies differ from the observed QPO pairs by less than $0.3$~Hz (Table~\ref{tab:evidence}).

\emph{Schwarzschild model.} The Schwarzschild model does not work for GRO J1655$-$40 and XTE J1550$-$564. The best fit goes to the lower boundary of the mass prior, and even there both QPO frequencies are missed by $\simeq25$--$30$~Hz, with $\chi^2_{\min}\simeq111$--$113$, $\Delta{\rm AIC}\simeq110$ and $\ln B\simeq-57$ relative to the best model. This is related to the known fact that for $A=1$ the RPM gives exactly the two QPO frequencies only for $M=4.09M_\odot$ and $M=6.68M_\odot$ (in both cases at $r=6.75M$), which is much lower than the dynamical masses. The mass of GRS 1915$+$105 is known with much lower precision, and for this source the Schwarzschild model fits the data well ($\chi^2_{\min}=0.48$ at $M\simeq11.1M_\odot$). It is even preferred by the information criteria and by the evidence, but the preference over the other models is inconclusive ($|\ln B|\le1.1$).

\emph{Letelier--KR geometry without magnetic field.} If there is no magnetic field, the only new parameter is $A$. For neutral particles the RPM equations can be solved analytically. From Eq.~\eqref{eq:Om0} we have $A-6M/r=(1-\nu_{\rm L}/\nu_{\rm U})^2$, and then for a given mass the needed value of $A$ is found from $\nu_{\rm U}$. For the central values of the masses we obtain $A=1.268$ for GRO J1655$-$40, $A=1.306$ for XTE J1550$-$564 and $A=1.104$ for GRS 1915$+$105, with emission radii $r\simeq1.10\,r_{\rm ISCO}$. However, for $\alpha\ge0$ and $\ell\le0.19$ we have $A\le1/(1-0.19)=1.235$. As a result, the MCMC gives $A=1.176^{+0.027}_{-0.026}$ for GRO J1655$-$40 and $A=1.209^{+0.013}_{-0.018}$ for XTE J1550$-$564; in these fits $\ell$ goes close to its upper limit ($\ell\simeq0.17$--$0.18$), $\alpha\simeq0$, and the masses are $\simeq1.4$--$1.6\sigma$ lower than the dynamical values (Table~\ref{tab:results}). The fit is acceptable for GRO J1655$-$40 ($\chi^2_{\min}=0.33$) and marginal for XTE J1550$-$564 ($\chi^2_{\min}=2.48$). Compared with the magnetized model, the evidence disfavours this model moderately for GRO J1655$-$40 ($\ln B=-3.2$) and strongly for XTE J1550$-$564 ($\ln B=-6.1$). For GRS 1915$+$105 the posterior of $A$ is consistent with one, $A=0.990^{+0.083}_{-0.079}$. For all sources, the data constrain only $A$, and the separate posteriors of $\alpha$ and $\ell$ follow from the priors along the line $A={\rm const}$. This can be seen from the strong $\alpha$--$\ell$ correlation in the corner plots.


\emph{Magnetized models.} A weak Larmor-type magnetic coupling allows us to fit the QPO pairs of GRO J1655$-$40 and XTE J1550$-$564 at the dynamical masses. For the magnetized Schwarzschild model we get $\beta=-0.0177^{+0.0035}_{-0.0034}$ and $\beta=-0.0199^{+0.0024}_{-0.0025}$, and the emission radii $r\simeq6.3$--$6.4M$ are a little outside the corresponding ISCO ($r_{\rm ISCO}\simeq5.6M$). For both sources, this model has the largest evidence. If the particles are protons, then $\beta=qBGM/(2m_pc^3)$ corresponds to $B\simeq0.14$~G for GRO J1655$-$40 and $B\simeq0.09$~G for XTE J1550$-$564. For electrons, the fields should be smaller by the factor $m_e/m_p$, i.e., $B\sim10^{-4}$~G. As was noted in Refs.~\cite{Kolos:2015iva,Tursunov:2016dss}, such weak fields still strongly affect charged particles because their specific charge is large. For these two sources, the anti-Larmor mode ($\beta>0$) has only $5.7\%$ and $0.5\%$ of the posterior mass. For GRS 1915$+$105 it has $33\%$, and the magnetic coupling of this source is consistent with zero, $\beta=-0.0075^{+0.0255}_{-0.0090}$ in the full model. When we add $(\alpha,\ell)$ to the magnetized model (the full model), the values of $\beta$ and $M$ and the quality of the fit almost do not change ($\ln B=-0.18$ and $-0.11$ for GRO J1655$-$40 and XTE J1550$-$564), and the posterior of $A$ stays close to its prior, $A\simeq0.93\pm0.11$ (Fig.~\ref{fig:Apost}). Thus, if the magnetic field is taken into account, the present QPO data do not require the CoS and the KR deformation, but they also do not exclude them. The penalty $\Delta{\rm AIC}=4$ of the full model comes only from its two extra parameters.

\begin{table*}[t]
\caption{Marginalized posterior constraints (median and $16$th/$84$th percentiles) for the four models and the three sources; $A=(1-\alpha)/(1-\ell)$ is a derived parameter. The last columns give the minimum $\chi^2$ (including the mass datum), $\Delta$AIC and $\Delta$BIC with respect to the best model of each source.}
\label{tab:results}
\centering\footnotesize\setlength{\tabcolsep}{3pt}
\begin{adjustbox}{max width=\textwidth}
\begin{tabular}{llccccccccc}
\toprule
Source & Model & $M/M_\odot$ & $\alpha$ & $\ell$ & $A$ & $\beta$ & $r/M$ & $\chi^2_{\min}$ & $\Delta$AIC & $\Delta$BIC \\
\midrule
GRO J1655$-$40 & (a) Schwarzschild & $4.80^{+0.01}_{-0.00}$ & -- & -- & -- & -- & $6.30^{+0.02}_{-0.02}$ & 112.54 & 110.54 & 111.44 \\
 & (b) $\beta=0$ & $4.98^{+0.14}_{-0.12}$ & $0.017^{+0.021}_{-0.012}$ & $0.170^{+0.014}_{-0.021}$ & $1.176^{+0.027}_{-0.026}$ & -- & $5.62^{+0.13}_{-0.14}$ & 0.33 & 2.33 & 1.43 \\
 & (c) $\alpha=\ell=0$ & $5.39^{+0.29}_{-0.28}$ & -- & -- & -- & $-0.0177^{+0.0035}_{-0.0034}$ & $6.38^{+0.12}_{-0.13}$ & 0.00 & 0.00 & 0.00 \\
 & (d) full & $5.40^{+0.28}_{-0.29}$ & $0.074^{+0.079}_{-0.051}$ & $0.021^{+0.096}_{-0.100}$ & $0.934^{+0.120}_{-0.111}$ & $-0.0178^{+0.0038}_{-0.0032}$ & $6.63^{+0.49}_{-0.52}$ & 0.00 & 4.00 & 2.20 \\
\midrule
XTE J1550$-$564 & (a) Schwarzschild & $8.51^{+0.02}_{-0.01}$ & -- & -- & -- & -- & $6.15^{+0.02}_{-0.02}$ & 111.26 & 109.26 & 110.16 \\
 & (b) $\beta=0$ & $8.63^{+0.14}_{-0.09}$ & $0.008^{+0.011}_{-0.006}$ & $0.182^{+0.006}_{-0.010}$ & $1.209^{+0.013}_{-0.018}$ & -- & $5.38^{+0.07}_{-0.06}$ & 2.48 & 4.48 & 3.58 \\
 & (c) $\alpha=\ell=0$ & $9.10^{+0.28}_{-0.28}$ & -- & -- & -- & $-0.0199^{+0.0024}_{-0.0025}$ & $6.31^{+0.09}_{-0.09}$ & 0.00 & 0.00 & 0.00 \\
 & (d) full & $9.11^{+0.28}_{-0.28}$ & $0.080^{+0.078}_{-0.055}$ & $0.013^{+0.092}_{-0.098}$ & $0.921^{+0.114}_{-0.106}$ & $-0.0199^{+0.0024}_{-0.0023}$ & $6.61^{+0.45}_{-0.44}$ & 0.00 & 4.00 & 2.20 \\
\midrule
GRS 1915$+$105 & (a) Schwarzschild & $11.11^{+0.47}_{-0.47}$ & -- & -- & -- & -- & $6.70^{+0.16}_{-0.14}$ & 0.48 & 0.00 & 0.00 \\
 & (b) $\beta=0$ & $10.99^{+1.10}_{-1.03}$ & $0.058^{+0.059}_{-0.040}$ & $0.059^{+0.070}_{-0.077}$ & $0.990^{+0.083}_{-0.079}$ & -- & $6.77^{+0.65}_{-0.57}$ & 0.00 & 3.52 & 1.72 \\
 & (c) $\alpha=\ell=0$ & $11.89^{+1.79}_{-0.96}$ & -- & -- & -- & $-0.0016^{+0.0232}_{-0.0126}$ & $6.47^{+0.28}_{-0.76}$ & 0.00 & 1.52 & 0.62 \\
 & (d) full & $12.31^{+1.80}_{-1.55}$ & $0.072^{+0.072}_{-0.049}$ & $0.026^{+0.083}_{-0.092}$ & $0.944^{+0.097}_{-0.098}$ & $-0.0075^{+0.0255}_{-0.0090}$ & $6.68^{+0.68}_{-0.85}$ & 0.00 & 5.52 & 2.82 \\
\bottomrule
\end{tabular}
\end{adjustbox}
\end{table*}
\begin{table}[t]
\caption{Bayesian evidences $\ln Z$ (from \texttt{dynesty}, statistical uncertainty $\simeq0.1$) and Bayes factors $\ln B=\ln Z-\ln Z_{\rm best}$ with respect to the best model of each source, together with the best-fit QPO frequencies of the full model.}
\label{tab:evidence}
\centering\footnotesize\setlength{\tabcolsep}{3.5pt}
\begin{adjustbox}{max width=\columnwidth}
\begin{tabular}{lccc}
\toprule
Model & GRO J1655$-$40 & XTE J1550$-$564 & GRS 1915$+$105 \\
\midrule
Schwarzschild & $-70.02\ (-57.43)$ & $-68.18\ (-56.41)$ & $-9.54\ (+0.00)$ \\
$\beta=0$ & $-15.79\ (-3.20)$ & $-17.87\ (-6.10)$ & $-10.27\ (-0.73)$ \\
$\alpha=\ell=0$ & $-12.59\ (+0.00)$ & $-11.77\ (+0.00)$ & $-10.28\ (-0.74)$ \\
full & $-12.78\ (-0.18)$ & $-11.88\ (-0.11)$ & $-10.63\ (-1.09)$ \\
\midrule
$\nu_{\rm U}^{\rm obs}/\nu_{\rm U}^{\rm th}$ [Hz] & 450/450.11 & 276/276.14 & 168/167.99 \\
$\nu_{\rm L}^{\rm obs}/\nu_{\rm L}^{\rm th}$ [Hz] & 300/300.00 & 184/183.97 & 113/112.97 \\
\bottomrule
\end{tabular}
\end{adjustbox}
\end{table}
\begin{table}[t]
\caption{Convergence diagnostics of the full-model \texttt{emcee} chains in the two sub-domains $\beta<0$ and $\beta>0$. $w$ is the posterior weight of each mode obtained with \texttt{dynesty}, $\tau_{\rm int}$ the range of integrated autocorrelation times over the five parameters, $N_{\rm step}$ the number of steps per walker and $N_{\rm s}$ the number of retained samples.}
\label{tab:conv}
\centering\footnotesize\setlength{\tabcolsep}{3pt}
\begin{adjustbox}{max width=\columnwidth}
\begin{tabular}{llccccc}
\toprule
Source & mode & $w$ & acc. & $\tau_{\rm int}$ & $N_{\rm step}/\tau_{\max}$ & $N_{\rm s}$ \\
\midrule
GRO J1655$-$40 & $\beta<0$ & 0.943 & 0.456 & 94--182 & 329 & 53150 \\
 & $\beta>0$ & 0.057 & 0.234 & 439--1629 & 123 & 38050 \\
XTE J1550$-$564 & $\beta<0$ & 0.995 & 0.494 & 74--97 & 620 & 69400 \\
 & $\beta>0$ & 0.005 & 0.333 & 200--501 & 120 & 25000 \\
GRS 1915$+$105 & $\beta<0$ & 0.674 & 0.478 & 86--96 & 1041 & 99200 \\
 & $\beta>0$ & 0.326 & 0.295 & 221--1291 & 77 & 37850 \\
\bottomrule
\end{tabular}
\end{adjustbox}
\end{table}

\section{Conclusion}\label{sec6}

In this work, we have studied the motion of charged test particles and the related quasi-periodic oscillations around the Letelier black hole in Kalb--Ramond gravity in the presence of an external magnetic field. The main results of the paper are the following.

\begin{enumerate}
	\item The CoS parameter $\alpha$ and the Lorentz-violating KR parameter $\ell$ appear in the metric only through the combination $A=(1-\alpha)/(1-\ell)$. The horizon, the photon sphere and the ISCO of neutral particles are located at $r_{\rm h}=2M/A$, $r_{\rm ph}=3M/A$ and $r_{\rm ISCO}=6M/A$. For $A\neq1$ the spacetime is not asymptotically flat, and the observed frequencies are rescaled by the factor $1/\sqrt A$. When $\alpha=\ell$, the effects of the CoS and the KR field cancel each other and we get the Schwarzschild geometry.

	\item The spacetime is not Ricci flat, so the Wald potential does not satisfy the Maxwell equations. We have obtained the exact regular source-free magnetic test field in closed form, $\Psi\propto{}_2F_1(s_+-1,-s_+;1;1-Ar/2M)$. At large distances it behaves as $r^{s_+}$ with $s_+=\frac12(1+\sqrt{1+8/A})$, and for $A=1$ it reduces to the Wald profile $r^2$.

	\item We have derived the equations of motion and the effective potential, and obtained closed-form expressions for the circular orbits and for the orbital, radial and vertical epicyclic frequencies. These expressions were checked by numerical integration of the full equations of motion. The magnetic field shifts the ISCO inward for both directions of the Lorentz force, the CoS shifts it outward and a positive KR parameter shifts it inward. The trajectories of charged particles (escaping, captured, trapped and curled) are sensitive to $A$, and $A$ can change the topology of the allowed regions.

	\item We have performed the MCMC analysis of the twin-peak QPOs of GRO J1655$-$40, XTE J1550$-$564 and GRS 1915$+$105, together with nested sampling for the Bayesian evidence and the weights of the two magnetic modes. We found that the Schwarzschild model is excluded for GRO J1655$-$40 and XTE J1550$-$564. The purely geometric deformation can explain the data only for $A\simeq1.18$--$1.21$; this needs $\ell$ close to the upper bound $0.19$ allowed by the current KR constraints and masses somewhat lower than the dynamical ones, and the evidence disfavours it moderately or strongly. A weak Larmor-type magnetic coupling $\beta\simeq-0.018$ to $-0.020$ fits both sources at their dynamical masses and has the highest evidence. For GRS 1915$+$105 all four models are consistent with the data.
\end{enumerate}

The main limitation of our results, which is common also for other similar studies, is the degeneracy between $\alpha$ and $\ell$. Only $A$ enters the test-particle dynamics, and therefore QPOs, observables related to the ISCO or other tests of the geometry alone cannot separate the CoS from the KR field. To break this degeneracy one needs observables which couple to the two sources in a different way, for example to the KR field itself via non-minimal couplings, or independent bounds on $\ell$ from Solar-System and laboratory experiments~\cite{Yang:2023wtu,Junior:2024ety}. Also, the number of free parameters is larger than the number of QPO data for each source, so the constraints depend on the chosen priors and on the RPM; for this reason we have listed the priors explicitly. Our analysis is based on the test-field and test-particle approximations, and we do not take into account the black-hole spin, the plasma dynamics and the back-reaction of the magnetic field. In future works it would be interesting to consider rotating Letelier--KR black holes, other QPO models such as the epicyclic resonance models~\cite{Abramowicz:2001bi}, sources with independent spin measurements~\cite{Motta:2013wga}, the intermediate-mass candidate M82 X-1~\cite{Pasham:2014ybe}, and to combine QPO data with shadow observations~\cite{AHMED2026102315}.

\section*{Data availability}
The Python codes used to compute the magnetic profile, the frequencies, the trajectories and the MCMC/nested-sampling posteriors are available from the authors upon reasonable request.

\appendix
\section{Exact regular magnetic profile}\label{app:A}

Here we present the derivation of Eq.~\eqref{eq:Psi}. Using $z=r/r_{\rm h}$ and $f=A(1-1/z)$, we can rewrite Eq.~\eqref{eq:maxwell} as
\begin{equation}
	\frac{d}{dz}\left[\Big(1-\frac1z\Big)\frac{d\Psi}{dz}\right]-\frac{2}{A}\frac{\Psi}{z^2}=0 ,
\end{equation}
or, equivalently,
\begin{equation}
	z(1-z)\Psi''-\Psi'+\frac{2}{A}\Psi=0 ,
	\label{eq:A1}
\end{equation}
where the prime now means $d/dz$. This is the Gauss hypergeometric equation $x(1-x)y''+[c-(a+b+1)x]y'-ab\,y=0$ with $x=z$, $c=-1$, $a+b=-1$ and $ab=-2/A$, so that $\{a,b\}=\{s_+-1,\,-s_+\}$, where $s_+$ is given by Eq.~\eqref{eq:spm}. Eq.~\eqref{eq:A1} has singular points at $z=0$ (curvature singularity), $z=1$ (horizon) and $z=\infty$. Near $z=1$ the exponents are $0$ and $c-a-b=0$. Hence, one of the solutions is analytic at the horizon, and the other one has a logarithmic divergence. Written in terms of $1-z$, the analytic solution is~\cite{Olver:2010dlmf}
\begin{align}
	\Psi_{\rm reg}(z)&={}_2F_1(a,b;a+b+1-c;1-z)\nonumber\\
	&={}_2F_1(s_+-1,-s_+;1;1-z),
\end{align}
and this is Eq.~\eqref{eq:Psi}. Its behaviour at $z\to\infty$ can be obtained from the connection formula for ${}_2F_1(a,b;c';x)$ at $x\to-\infty$,
\begin{align}
	{}_2F_1(a,b;c';x)\simeq{}&\frac{\Gamma(c')\Gamma(b-a)}{\Gamma(b)\Gamma(c'-a)}(-x)^{-a}\nonumber\\
	&+\frac{\Gamma(c')\Gamma(a-b)}{\Gamma(a)\Gamma(c'-b)}(-x)^{-b},
\end{align}
For $c'=1$, $-b=s_+$ and $-a=1-s_+=s_-$ this gives $\Psi_{\rm reg}\simeq C_+z^{s_+}+C_-z^{s_-}$ with $C_+=\Gamma(2s_+-1)/[\Gamma(s_+-1)\Gamma(s_++1)]$, and we arrive at the normalization~\eqref{eq:norm}. In the case $A=1$ we have $a=1$, $b=-2$, the series is finite and ${}_2F_1(1,-2;1;1-z)=z^2$. The derivatives used in Eqs.~\eqref{eq:Xsol}--\eqref{eq:Vpp} are
\begin{align}
	\Psi'(r)&=-\frac{\mathcal N}{r_{\rm h}}\,ab\;{}_2F_1(a+1,b+1;2;1-z),\\
	\Psi''(r)&=\frac{\mathcal N}{r_{\rm h}^2}\,\frac{a(a+1)b(b+1)}{2}\;{}_2F_1(a+2,b+2;3;1-z).
\end{align}
All numerical results of this paper are obtained with these closed-form expressions, which we checked by high-precision numerical differentiation (the relative difference is below $10^{-15}$).

\bibliographystyle{apsrev4-2}
\bibliography{Refernces}
\end{document}